# Reflection, emission, and polarization properties of surfaces made of hyperfine grains, and implications for the nature of primitive small bodies

Robin Sultana[1,2], Olivier Poch[1], Pierre Beck[1], Bernard Schmitt[1], Eric Quirico[1], Stefano Spadaccia[3], Lucas Patty[3], Antoine Pommerol[3], Alessandro Maturilli[4], Jörn Helbert[4], Giulia Alemanno[4]

[1]Univ. Grenoble Alpes, CNRS, IPAG, 38000 Grenoble, France

[2]now at LATMOS, 11 Boulevard d'Alembert 78280 Guyancourt, France

[3]Physikalisches Institut, University of Bern, Switzerland, NCCR PlanetS.

[4]DLR, Institute of Planetary Research, Berlin, Germany

## Highlights

- We investigate mixtures of hyperfine powders of bright silicates and opaque minerals
- Grain separation strongly impacts the optical properties of these mixtures
- Spectral bluing occurs when opaque grains are dispersed in a matrix of diffusing grains
- A 10-µm emissivity signature is observed for hyperfine silicates mixed with opaque grains
- Polarimetric properties show a non-linear behaviour of the mixture

## Abstract

Solar System small bodies were the first objects to accrete inside the protoplanetary disk, giving insights into its composition and structure. The P-/D-type asteroids are particularly interesting because of the similarity of their spectra, at visible and near infrared wavelengths (Vis-NIR), with cometary nuclei, suggesting that they are the most primitive types of small bodies. There are various indications that (1) their low albedo in the visible (Vis) and mid-infrared (MIR) wavelength ranges seems mainly controlled by the presence of opaque minerals (iron sulfides, Fe-Ni alloys etc.) (Quirico et al., 2016; Rousseau et al., 2018); and (2) their surfaces are made of intimate mixtures of these opaque minerals and other components (silicates, carbonaceous compounds, etc.) in the form of sub-micrometre-sized grains, smaller than the wavelength at which they are observed, so-called "hyperfine" grains. Here, we investigate how the Vis-NIR-MIR (0.55-25 µm) spectral and V-band (0.53 µm) polarimetric properties of surfaces made of hyperfine grains are influenced by the relative abundance of such hyperfine materials, having strongly different optical indexes. Mixtures of grains of olivine and iron sulfide (or anthracite), as analogues of silicates and opaque minerals present on small bodies, were prepared at different proportions. The measurements reveal that these mixtures of hyperfine grains have spectral and polarimetric Vis-NIR properties varying in strongly nonlinear ways. When present at even a few percent, opaque components dominate the Vis-NIR spectral and polarimetric properties, and mask the silicate bands at these wavelengths. The Vis-NIR spectral slope ranges from red (positive slope), for pure opaque material, to blue (negative slope) as the proportion of silicates increases, which is reminiscent of the range of spectral slopes observed on P/D/X/C- and B-types asteroids. The spectra of the darkest mixtures in the Vis-NIR exhibit the absorption bands of Si-O in olivine around 10 µm in the MIR, which is observed in emission for several small bodies. The samples studied here have macro- and micro-porosities lower than 78%, indicating that surfaces more compact than "fairy castle" hyperporous (80-99%) ones can also exhibit a blue spectral slope or a silicate signature at 10 µm. Remarkably, some mixtures exhibit altogether a red spectral slope in the Vis-NIR, a 10-µm feature in the MIR, and a V-band polarimetric phase curve similar (but not identical) to P-/D-type asteroids, reinforcing the hypothesis that these bodies are made of powdery mixtures of sub-micrometre-sized grains having contrasted optical indexes. This work shows that both the contrasted optical indexes of the components, and the dispersion or aggregation −depending on their relative proportions− of their hyperfine grains, induce different light scattering regimes in the Vis-NIR and MIR, as observed for primitive small bodies. The optical separation of hyperfine grains seems to be a major parameter controlling the optical properties of these objects.

## 1. Introduction

*1.1. Small bodies reflection, emission, and polarization properties*

Our Solar System contains a variety of small body populations, which date from the first phases of its formation. A significant fraction of these objects did not accumulate enough mass, or not fast enough, to achieve temperature sufficient to induce partial or full differentiation. Hence, they hold key information about the materials and the conditions inside the protoplanetary disk where they formed. During the early stage of planet formation, migrations of the giant gaseous planets should have disturbed the orbit of small bodies (Morbidelli et al., 2005; Tsiganis et al., 2005), which complexifies the retrieval of protoplanetary disk compositional and thermal structure. Still, large optical surveys of small bodies revealed a gradient in composition with an evolution in spectral types throughout the asteroid main belt, with the dominant types being successively S, C and P/D- types with increasing heliocentric distance (Gradie and Tedesco, 1982; DeMeo and Carry, 2013).

The P/D-type asteroids, found in the main belt and among Jupiter Trojans, are particularly interesting because their visible to mid-infrared spectra are similar to comets, suggesting that they are the most primitive types of asteroids (Vernazza and Beck, 2017). Reflectance spectra in the visible and near infrared (Vis-NIR) of cometary nuclei and P/D-type asteroids both display similar red to flattish spectral slopes, lacking absorption features (Emery et al., 2011; Capaccioni et al., 2015). In the 3-µm spectral region, absorption bands possibly due to $NH_4^+$ are present in P/D-types, and comet 67P/Churyumov-Gerasimenko (Poch et al., 2020). In the mid-infrared (MIR) range, emission spectra of cometary comae and P/D-type objects display a similar emissivity plateau from around 9 to 11 µm caused by the fundamental mode of vibrations of Si-O in silicates (Emery et al., 2006; Vernazza et al., 2012). A similar but not identical and generally weaker emissivity feature is observed on C-type and on some X-type asteroids (Emery et al., 2011; Marchis et al., 2012; Vernazza et al., 2017), and is more contrasted for larger objects (Marchis et al., 2012). On the contrary, B-type asteroids such as Pallas, Phaethon and Bennu, do not exhibit an emissivity plateau around 10 µm (Lim et al., 2005; McAdam et al., 2018; Lim et al., 2019; Hamilton et al., 2019b). In term of albedo and Vis-NIR color, these C/X/B-type asteroids are part of the same cluster of main-belt asteroids (Beck and Poch, 2021). Their Vis-NIR spectral slope span from slightly red (X/C-type) to significantly blue (B/C-type), and is redder for larger objects (Beck and Poch, 2021). Each of these spectral characteristics have been independently attributed to the surface texture of the objects, smaller asteroids being covered by rocks or large dust particles, while larger ones are covered by finer particles (Marchis et al., 2012; Beck and Poch, 2021). This apparent correlation between Vis-NIR slope and 10-µm feature is also observed for Trojan asteroids (D-type) (Emery et al., 2011). In addition, low-albedo asteroids exhibit differences in the degree of linear polarization of the visible light they scatter at different phase angles, i.e. their polarimetric phase curves. Absolute values of the minimum of polarization in the negative branch of the polarimetric phase curves (hereafter called $P_{min}$) and inversion angles (hereafter called $α_{inv}$) of F/P/D-type asteroids are generally smaller compared to the values found for B/C/Ch-type asteroids (Belskaya et al., 2017, 2019) (except for P-type for which $\alpha_{inv}$ is similar to B-type). This has also been tentatively attributed to differences of microstructure of their regolith, F/D-types having an optically more homogenous microstructure (Belskaya et al., 2005; Bagnulo et al., 2015). Taken together, these observations suggest a possible correlation between the redness of Vis-NIR spectral slope, the presence of a contrasted 10-µm plateau, and the absolute values of the polarimetric parameters $P_{min}$ and possibly $\alpha_{inv}$. However, to our knowledge, the way these polarimetric and spectral properties in the Vis-NIR and MIR vary for specific asteroidal surfaces analogues of controlled microstructure has never been tested numerically nor experimentally.

These observations give rise to the following questions: Why are P/D-type asteroids spectrally similar to comets, in particular in the MIR where the emissivity spectra of asteroidal surfaces are similar to that of comae's clouds of particles? Among the P/D/X/C- and B-type asteroids, what cause(s) the Vis polarisation phase curve, Vis-NIR spectral slope and the 10-µm MIR variations? As suggested by their possible correlation, could the variations of spectral and polarimetric features be due to similar textural properties of these small bodies' surfaces?

We now have multiple measurements − of inter-planetary dust particles collected on Earth (Rietmeijer, 1993), of samples returned from comets (Price et al., 2010), or analysed in situ (Mannel et al., 2019) −, all indicating that the surface material of comets and possibly C-, P- and D-types asteroids (Vernazza et al., 2015) is made of agglomerates of individual sub-micrometric monomers. The dust agglomerates observed at comet 67P/Churyumov-Gerasimenko have different morphologies from compact (porosity <10%) to porous (10–95%) or fluffy (> 95%) particles (Güttler et al., 2019). As in the case of interplanetary dust particles (IDPs), micro-meteorites, or primitive meteorites matrices, the different constituents (silicates, opaque minerals, carbonaceous compounds etc.) are mixed at the sub-micrometric scale. As discussed and shown in Quirico et al. (2016) and Rousseau et al. (2018), the low albedo of these cosmo-materials from the Vis to the MIR wavelength range seems mainly controlled by the presence of opaque minerals (iron sulfides, Fe-Ni alloys etc.) rather than by other major components such as silicates and carbonaceous materials. Strikingly, the spectra of comets, P/D-type asteroids and Jupiter Trojans are perfectly matched by spectra of anhydrous chondritic porous IDPs (CP-IDPs) (see Figure 1 in Vernazza et al., 2015), which are predominantly made of aggregates of sub-micrometer-sized grains of silicates (~20-50 vol%), Fe-Ni sulfides (≤ 40 vol%) and carbonaceous materials (≤ 40 vol%), arranged in a fluffy microstructure (Alexander et al., 2007; Bradley, 2014). Therefore, CP-IDPs appear as convincing analogues of the surface of comets and primitive asteroids, but how the various compositional and textural (grain size, porosity) parameters influence the surface optical properties is still an open question.

In the following paragraph, we review the current knowledge of the MIR, Vis-NIR and polarimetric properties of particulate surfaces, especially those made of grains smaller than 1 µm and of dark or mixed bright/dark components.

*1.2. Interpretations of small bodies reflection and emission properties*

Typical MIR spectra of silicate minerals are characterized by reststrahlen bands (corresponding to the Si-O fundamental modes at 9-11 µm, and 18-28 µm), a Christiansen feature (where the refractive index varies rapidly with wavelength, near 8 µm for most silicates), and transparency features between the reststrahlen bands (Mustard and Glotch, 2020). A large number of studies have shown how grain size (Hunt and Vincent, 1968; Hunt and Logan, 1972; Arnold and Wagner, 1988; Moersch and Christensen, 1995; Mustard and Hays, 1997; Le Bras and Erard, 2003) and surface porosity (Salisbury and Eastes, 1985; Salisbury and Wald, 1992; Salisbury et al., 1994) influence these mid-infrared spectral bands and features. Most of these measurements showed that mid-infrared spectral features progressively vanish with decreasing grain size from few hundreds to few micrometres. However, for grains smaller than 5 µm, Hunt and Logan (1972) and Arnold and Wagner (1988) observed an increased spectral contrast around 10 µm, with emission maxima. The work of Salisbury and Eastes (1985) shows that the spectral contrast of such small grains is very dependent on packing conditions, i.e. the surface porosity. Based on these laboratory measurements and on a Hapke-Mie hydride radiative transfer model, Emery et al. (2006) proposed that the emission feature observed between 9 and 11 µm for Trojan asteroids is due to very fine silicate grains either in a very porous surface structure, or embedded in a more transparent material hypothesized to behave as void-space, which could explain why the same feature is observed for comae. Following this work, several studies have measured reflectance spectra of samples made by suspending minerals or meteorites powders in infrared-transparent potassium bromide (KBr) powder to simulate void spaces in the MIR. These spectra have mid-infrared spectral features similar to those observed on Trojan and main-belt asteroids. These results were interpreted as a confirmation of Emery et al. (2006) suggestions, i.e. either the presence of IR-transparent salts (King et al., 2011; Yang et al., 2013; Izawa et al., 2021), or the high porosity of asteroids surfaces (Vernazza et al., 2012; Young et al., 2019; Martin et al., 2022), the latter being consistent with their thermal inertia and radar albedo (Vernazza et al., 2012). The presence of clouds of grains in electrostatic levitation over the surface is also a possible explanation (Vernazza et al., 2012; Wang et al., 2016).

However, the mixtures reproducing the 10-µm feature are those containing a relatively large fraction of IR-transparent KBr salt grains used to simulate void spaces, which do not behave exactly as such in the radiative transfer. Indeed, these KBr grains scatter the light and make the samples relatively bright in the Vis-NIR and MIR reflected light (equivalent to a low emissivity in the MIR according to the Kirchoff's law), which is inconsistent with observations of low-albedo asteroids and comets (which have a low reflectance in the Vis-NIR and a high emissivity in the MIR). Using a modified Hapke model, Yang et al. (2013) showed that the Vis-NIR and MIR spectra of Trojan asteroids can both be explained by fine grained silicates (1–5 wt%) and highly absorbing material (2–10 wt%) suspended in a transparent matrix of IR-transparent salts. However, such a high abundance of these peculiar salts on asteroids and comets remains to be proven (e.g. the ammoniated salts detected on Ceres and comet 67P surfaces are absorbent in the MIR), and alternative explanations may be possible.

The possible origins of the Vis-NIR spectral slopes of small bodies, and their variability with grain size, porosity, mixtures, space weathering etc. have been the subject of many studies. A spectral reddening is observed after grinding minerals and meteorites (Ross et al., 1969; Johnson and Fanale, 1973; Cloutis et al., 2018; Beck et al., 2021), and a bluing is observed when sub-micrometre-sized or micrometre-sized grains of opaque minerals are present, embedded in mixtures with brighter grains (Clark et al., 2008; Loeffler and Prince, 2022) and/or in highly porous structures (Poch et al., 2016; Schröder et al., 2021). Bluing or reddening are also observed after laser or ion irradiations simulating space weathering (Lantz et al., 2017; Matsuoka et al., 2020). The explanations for these changes of Vis-NIR spectral slope are not always well-established, and, apart from changes of composition (after grinding, or irradiation), the changes of micro-structure and/or the appearance of sub-micrometre-sized scatterers are suspected to induce these effects (Beck and Poch, 2021).

However, very few systematic laboratory spectral measurements of such sub-micrometre-sized materials have been made. In a previous work (Sultana et al., 2021), we explored the influence on Vis-NIR reflectance spectra of the presence of sub-micrometre-sized grains, which are smaller than the wavelength at which they are observed, so called *hyperfine grains*. In this previous study, we only considered weakly absorbing materials (k < 0.01), and we showed that hyperfine grains have lower contrast of absorption bands, and lower spectral slopes than larger µm-size grains. We concluded that the presence of these hyperfine grains tends to uniformize the spectra of different material, driving them toward blue featureless spectra (*spectral degeneracy*), especially when they form hyper-porous structures, where they tend to scatter light as smaller aggregates of a few sub-µm grains.

In this work, with the goal to improve our knowledge on the causes explaining the optical properties of primitive low-albedo asteroids and comets, we investigate altogether the Vis-NIR and MIR spectral and polarimetric properties of surface mixtures of sub-micrometre-sized grains made of various proportions of strongly and weakly absorbing materials. Such mixtures are more relevant to low-albedo primitive bodies of the Solar system, and have some similarities with IDPs thought to come from these bodies. We obtained reflectance spectra in the wavelength range from 0.55 to 25 µm, that can also be used to infer emission properties of such samples, as well as polarization measurements at 0.53 µm. We compare our laboratory results to observations and discuss the implications in term of composition and texture of a few classes of Solar system small bodies.

## 2. Methodology

### 2.1 Samples preparation

Our goal was to produce heterogeneous aggregates of sub-µm grains made of two materials having contrasted optical indexes and mixed in different volume proportions, as *optical models* of the surface of primitive small bodies. The first material was the silicate mineral olivine, one the most abundant minerals in primitive meteorites and IDPs (Bradley, 2014; Scott and Krot, 2014) and detected in cometary particles (Brownlee, 2014). We used a Mg-rich forsterite olivine (Fo = [Mg] / ([Mg]+[Fe]) ≥ 94) purchased from Donghai Pellocci Crystal Products, China. For the second component, we chose iron sulfide and anthracite because they are absorbent over the Vis to MIR wavelength range, as the opaque minerals found in primitive meteorites and IDPs (mainly iron sulfides and Fe-Ni alloys, see Dai and Bradley, 2001; Quirico et al., 2016 and references herein). Moreover, we note that Fe-Ni sulfides are the second most common minerals in CP-IDPs after crystalline silicates (Dai and Bradley, 2001; Bradley, 2014). Iron sulfide was purchased from Alfa Aesar (ref. A15569.0B). X-ray diffraction of this powder indicates a composition made of 55 vol% troilite (FeS) and 45 vol% pyrrhotite ($Fe_{1-x}S$ with $0 < x < 0.2$). For simplicity, this sample will be labelled as "FeS" in the rest of the text. Anthracite was obtained from the Musée de la Mure (France). It is made of 91.9 % carbon, almost no oxygen atom, and is composed of graphite-like domains connected to less-organized polyaromatic matter (Albiniak et al., 1996). We stress that anthracite is not a material representative of the carbonaceous compounds expected on primitive small bodies (as discussed in Quirico et al., 2016), but we have chosen it here as another *optical analogue* of opaque minerals expected to be present on these bodies. The spectra of these endmember materials are presented in Figure 1 for hyperfine grains. Sulfide and anthracite display relatively flat spectra in the Vis-NIR and in the MIR. In the MIR, their spectra have a higher reflectance than olivine, and do not have strong spectral features (Fig. 1).

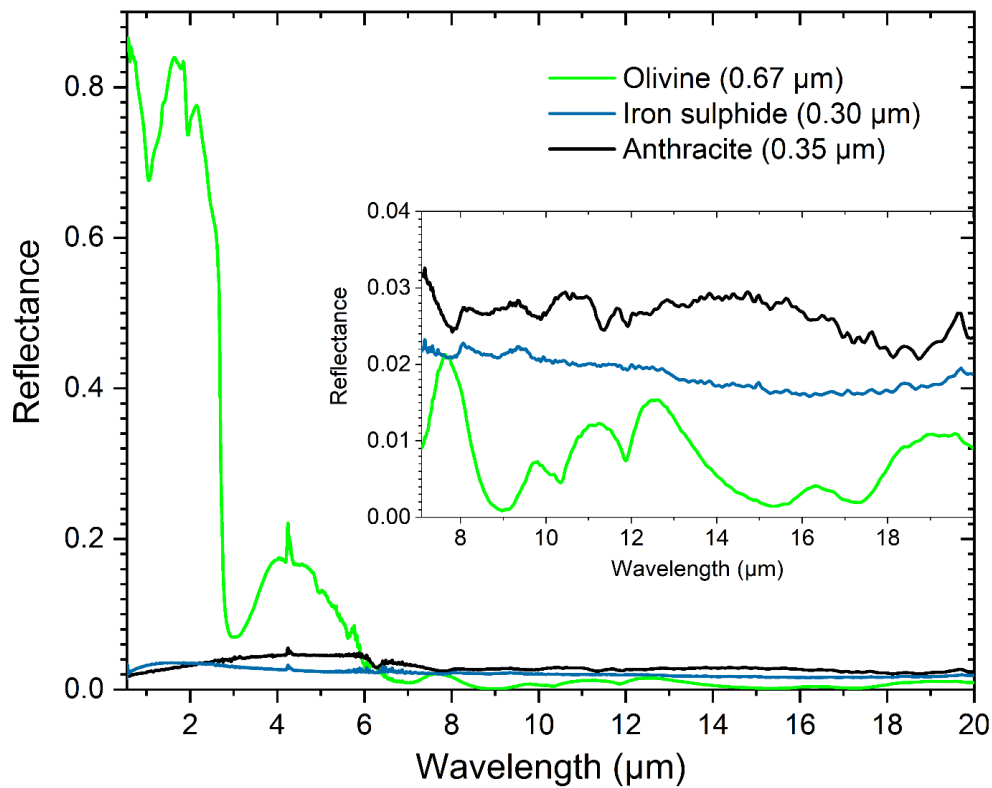

*Figure 1:* Reflectance in the Vis-NIR and in the MIR of olivine, iron sulfide and anthracite powders of sub-micrometric grain size. Spectra of iron sulfide and anthracite are relatively flat and do not present any absorption feature in the Vis-NIR or in the MIR. Note that the olivine is very bright in the Vis-NIR, but completely dark in the MIR. Iron sulfide and anthracite are both dark in the Vis-NIR but they are significantly brighter than olivine in the MIR. The small peak at 4.25 µm is related to gaseous $CO_2$.

To synthesize hyperfine powders of different materials, we used the grinding protocol described in detail in Sultana et al. (2021) to obtain grain diameters below 1 µm. Each sample is successively dry- (20 min) and wet-ground (150 min, in ethanol) several times using a Planetary Grinder Retsch© PM100 with progressively decreasing sizes of zirconium oxide ($ZrO_2$) grinding balls. After grinding, the balls are separated from the sub-micrometre-sized powder by sieving, and the ethanol is evaporated. Images obtained with a Scanning Electron

Microscope (SEM) were used to measure the grain size distribution of the obtained powders. The mean and maximal grain sizes in the powders for each component are listed in Table 1.

| Sample | Olivine | Iron sulfide | Anthracite |
| --- | --- | --- | --- |
| Mean grain size (µm) | 0.69 | 0.30 | 0.33 |
| Standard deviation (µm) | 0.47 | 0.28 | 0.35 |
| Median grain size (µm) | 0.56 | 0.21 | 0.21 |
| Max. grain size (µm) | 4.67 | 2.75 | 3.84 |

*Table 1: Mean, standard deviation, median and maximal sizes of the grains in each hyperfine powder prepared with the grinding protocol developed in Sultana et al. (2021). The size distributions are shown in Supplementary Figure 3.*

Mixing the two components (either iron sulfide or anthracite with olivine) was achieved by adding the two powders in a mortar and mixing them by hand grinding with the pestle for about 10 min. Powders were weighed separately to target the desired volume fraction of the mixture, using a Sartorius Quintix 35-1S scale, with a precision of 10 µg. To enable this preparation, the densities of the materials were measured using a pycnometer for the olivine (3.32 g.cm$^{-3}$) and iron sulfide (4.82 g.cm$^{-3}$) samples, and via the liquid displacement method for the anthracite (1.62 g.cm$^{-3}$). To ensure that the materials were well mixed together, they were imaged using a SEM. Figure 2 shows images of olivine and iron sulfide mixtures. These images are obtained in backscattered electron mode, sensitive to the mean molecular mass of the probed area. As the FeS is composed of atoms heavier than olivine, it backscatters electrons more efficiently, and FeS grains appear brighter in the images, whereas the olivine appears in lighter grey tone. Figure 2 shows how the grains of iron sulfide are efficiently dispersed in a matrix of olivine grains. The FeS grains do not form large aggregate and they are all separated by several olivine grains.

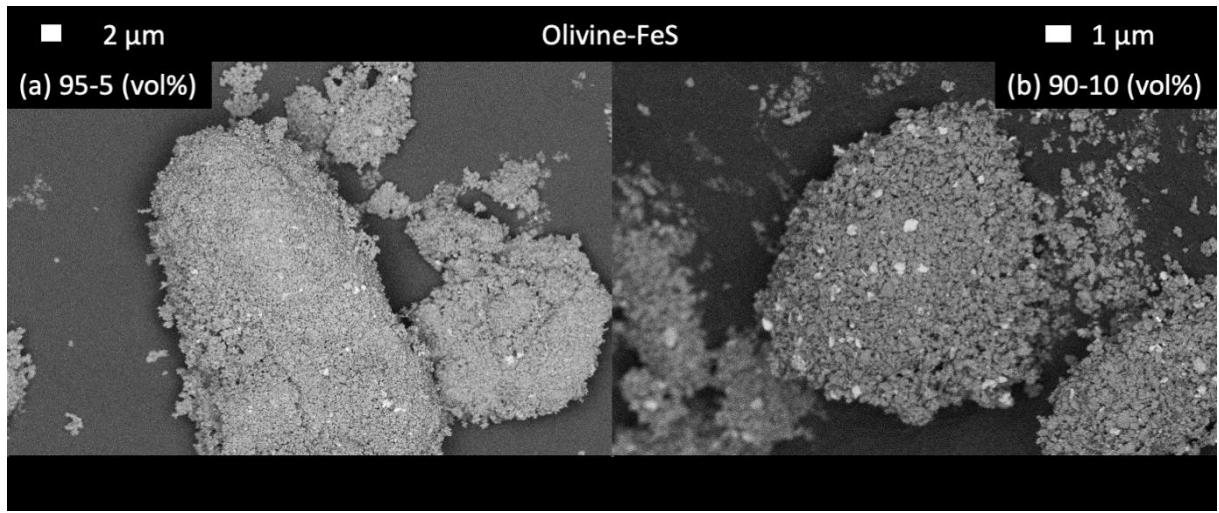

*Figure 2: Scanning Electron Microscope (SEM) images of mixtures of sub-µm grains of olivine and iron sulfide at different volume proportions. On this figure, we can observe that the grains of iron sulfide (brighter grains) are well dispersed in a matrix of olivine grains. They do not form aggregates of several grains of the same materials.*

## 2.2 Spectral and polarimetric measurements

Before each measurement, the powders were deposited into a sample holder (several millimetres thick and wide) using a spatula, and their surface was gently levelled to obtain a flat surface. Knowing the volume of the sample holders and the mass of powder introduced, we evaluated the samples porosity to range from 70% (for pure FeS powder) to 78% (for pure olivine powder). However, the porosity at the scale of the whole sample (several hundreds of µm up to mm, i.e. the "macro-porosity") may not be relevant for the light scattering that occurs at nm-µm scales. As seen on Figure 2, the sub-µm grains are assembled in individual aggregates several-micrometres-large (the largest aggregates seen on these images are 10 to 30 µm large), where they are in close contact with each other, and not in fluffy aggregates. Therefore, the bulk porosity (70-78%) might be mostly controlled by the void-spaces located between these multiple aggregates. More relevant for the light scattering is the surface rugosity and the sub-surface porosity at the nm-µm scales accessible to the photons (i.e. the "micro-rugosity" and "micro-porosity"). The SEM images (Figure 2) indicate that the micro-porosity of the samples studied here is possibly around ~50-70%, and definitely lower than that of fluffy aggregates obtained after sublimation of mixtures of water ice and mineral grains, measured in our previous study (see Figure 4 in Sultana et al., 2021; these samples had macro-porosity of 95-99%).

*Vis-NIR spectroscopy*: Reflectance spectra in the Vis-NIR range (0.55-4.2 µm) were obtained at IPAG with the spectro-gonio radiometer SHADOWS in standard mode (Potin et al., 2018; Sultana et al., 2021). The data were calibrated by dividing the sample signal by that of two reference targets under the same illumination-observation geometry. Vis-NIR spectra of each mixture and the pure materials are presented in Figure 3 and 4. Spectral resolution and sampling were similar to Sultana et al. (2021).

*Emissivity measurements*: Direct emissivity measurements from 5 to 17 µm were performed at the Planetary Spectroscopy Laboratory (PSL; Maturilli et al., 2019) at DLR in Berlin. The emissivity spectra were obtained by heating the sample, poured into aluminium sample holder, up to 673 K and measuring the emitted radiance with a Brucker Vertex 80V FT-IR spectrometer. The data were calibrated by dividing the sample radiance by the emitted radiance at the same temperature of a blast furnace slag, taken as blackbody, under the same temperature and observation conditions.

*MIR reflectance:* Studying the emissivity of mixtures was also performed by measuring the reflectance ($r_\lambda$) of the samples, and approximating the emissivity ($e_\lambda$) using Kirchhoff's thermal law: $e_\lambda = 1 - r_\lambda$. These MIR reflectance spectra from 1 to 25 µm were obtained at IPAG in Grenoble, using a Brucker Vertex 70V FT-IR spectrometer equipped with a biconical reflectance kit A513/QA. However, the Kirchhoff's law is only valid for hemispherical reflectance, and not for biconical reflectance, so the resulting absolute emissivity and spectral contrast are not reliable. The reflectance measurements were performed at ambient pressure and under dry air. Since the reflectance kit was not designed to study small phase angle and illuminates the sample at normal incidence, we used a 45º phase angle (i = 15º, e = 30º).

*Polarimetric measurements:* The polarimetric measurements were carried out on olivine, FeS and their mixtures with the POLarimeter for ICE Samples (POLICES) at the University of Bern (Poch et al., 2018; Spadaccia et al., 2022). A polarimeter (Hinds Dual PEM II/FS42-47) is placed above the sample at emission e = 0°, and the sample is illuminated by a motorized arm holding a collimating head fed by an optical fiber connected to a 530 nm LED (Thorlabs M530F2). The beam is depolarized and creates a 15 mm light spot on the sample at normal incidence. In this configuration, the incidence angle is equal to the phase angle, and it was varied by rotating the motorized arm from i = 1° to i = 70°. The linear polarization measurements are the result of the averaging of four measurements performed by rotating the sample on the horizontal plane with 45° incremental steps. This azimuthal averaging mitigates geometry effects due to tilts and orientations of the sample surface.

## 3. Results

### 3.1 Vis-NIR spectra of mixtures of hyperfine opaques grains and olivine

Figure 3 and Figure 4 present the evolution of olivine/FeS and olivine/anthracite mixtures with decreasing volume fraction of olivine. For both mixtures, we present the reflectance spectra (Fig. 3a, Fig. 4a) and spectra normalized at 0.7 µm (Fig. 3b, Fig. 4b). In both cases, we observe a general decrease of reflectance with increasing concentration of opaque grains. The olivine crystal-field absorption band at 1 µm becomes indistinguishable when the amount of opaque grains is in excess of 1 vol% for anthracite and 5 vol% for FeS. Interestingly, the Vis-NIR spectral slope from 0.5 to 1.5 µm is modified as the concentration of opaque increases: it becomes bluer for up to 10 vol% opaques, and redder for larger concentrations (Fig. 3) (by "bluer" or "redder", we mean that the relative spectral slope is getting more negative or positive, respectively, while the reflectance is decreasing at all wavelengths). At wavelengths larger than 1.5 µm, the spectral slope remains dominated by the spectral properties of olivine, FeS or anthracite depending on their proportions.

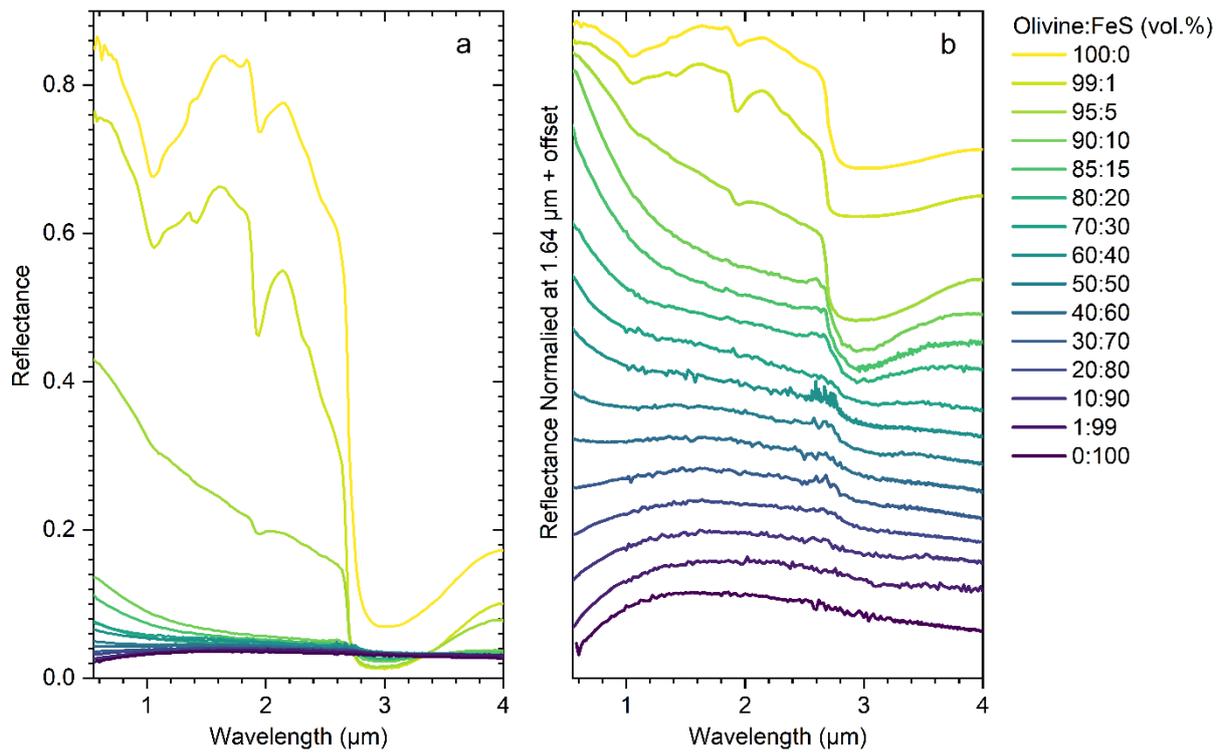

*Figure 3: Evolution of the reflectance spectra with increasing volume fraction of iron sulfide in the sample. The panel (**a**) shows the fast decrease of the reflectance level with the increasing amount of opaque. On the panel (**b**) the spectra are normalized at 1.64 µm and vertically shifted for the sake of clarity, so that the modifications of spectral slope are more visible. With the increasing fraction of opaque iron sulfide, the visible spectral slope becomes more and more blue, then neutral, and finally redder.*

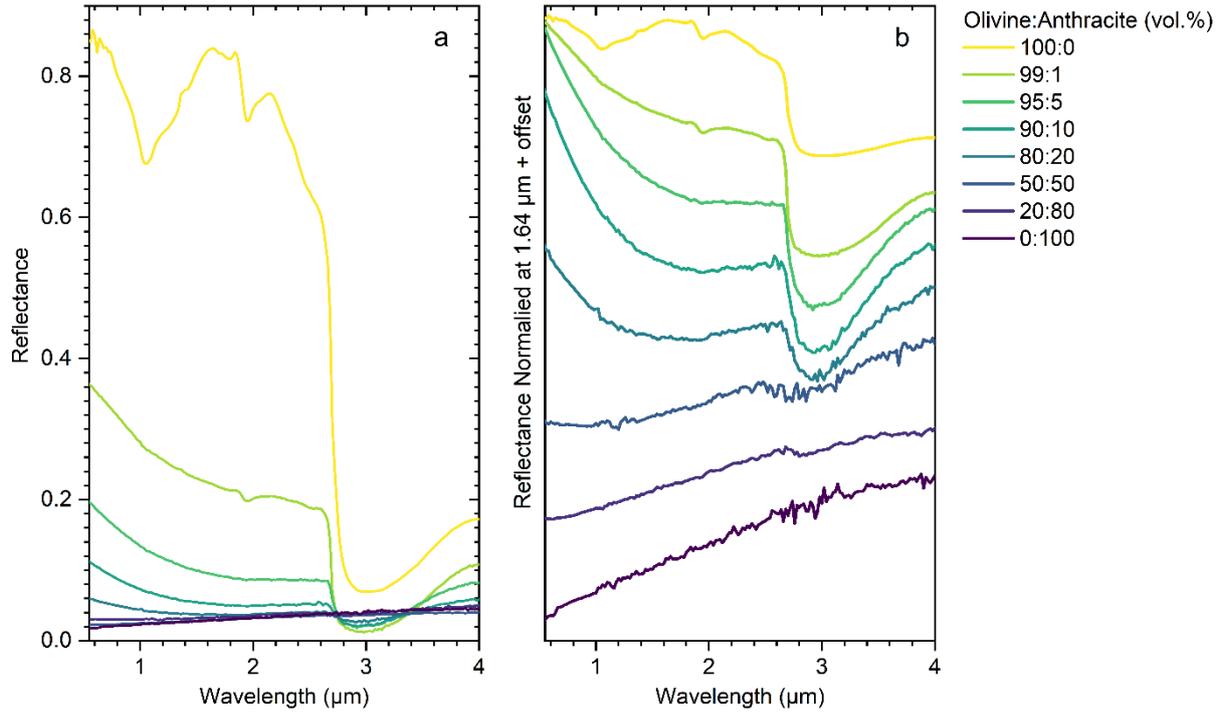

*Figure 4: Evolution of the reflectance spectra of hyperfine olivine powders with the increasing fraction of anthracite in the sample. The panel (**a**) shows the fast decrease of the reflectance level with the increasing amount of anthracite. On the panel (**b**) the spectra are normalized at 1.64 µm and shifted vertically to better display the modifications of spectral slope. With the increasing fraction of opaque anthracite, the visible spectral slope becomes more and more blue, and then redder.*

Spectral slopes were mostly calculated between 0.64 and 1.2 µm and defined as per cent of reddening per 100 nm (%.(100 nm)$^{-1}$) according to Delsanti et al. (2001) and Fornasier et al. (2015):

$$slope(\lambda_1, \lambda_2) = \frac{R_{\lambda_2} - R_{\lambda_1}}{R_{\lambda_1} \times (\lambda_2 - \lambda_1)} \times 10^4$$

Where λ is the wavelength in nm, and $R_{\lambda_1}$, $R_{\lambda_2}$ the absolute reflectance at $\lambda_1$, $\lambda_2$.

The computed values of the spectral slope and reflectance at 0.7 µm are provided in Table 2. For the highest fractions of olivine, the slope was computed on a different spectral interval (0.64-1.64 µm) to account for the presence of the $Fe^{2+}$ absorption band of olivine around 1 µm.

| VOLUME FRACTION Olivine-Opaque (%) | SPECTRAL SLOPE (%.(100nm)$^{-1}$) [0.64; 1.2] µm | | REFLECTANCE AT 0.7 µm (%) | |
|---|---|---|---|---|
| | Olivine-FeS | Olivine-Anthracite | Olivine-FeS | Olivine-Anthracite |
| 100-0 | -0.1* | -0.1* | 85.1 | 85.1 |
| 99-1 | -1.2* | -4.7 | 75.6 | 35.7 |
| 95-5 | -4.2* | -6.4 | 42.4 | 19 |
| 90-10 | -7.0 | -6.8 | 13.3 | 10.8 |
| 85-15 | -6.3 | - | 10.6 | - |
| 80-20 | -5.6 | -5.1 | 10.6 | 5.8 |
| 70-30 | -3.7 | - | 7.4 | - |
| 60-40 | -2.9 | - | 6.3 | - |
| 50-50 | -1.0 | -0.4 | 4.8 | 3.0 |
| 40-60 | 0.2 | - | 4.3 | - |
| 30-70 | 1.4 | - | 3.6 | - |
| 20-80 | 3.1 | - | 3.2 | - |
| 10-90 | 4.9 | - | 2.7 | - |
| 1-99 | 7.3 | - | 2.3 | - |
| 0-100 | 7.4 | 6.0 | 2.5 | 1.8 |

*Table 2: Values of spectral slopes and reflectance for the hyperfine olivine-opaque mixtures. *Due to the presence of the 1-µm absorption band of the olivine on these spectra, the interval to compute the spectral slope was changed to [0.64;1.64] µm.*

Figure 5a shows that the Vis-NIR slope decreases very fast to negative values (blue slope) with the amount of opaque, goes through a minimum value of about -7 %.100 nm$^{-1}$ for 10 vol% of opaque, then increases and reverts from blue to red around 60 vol% of opaque material, before reaching the slope of the pure opaque material spectrum.

In Figure 5b we present the evolution of reflectance with increasing amount of olivine in the mixture. The reflectance remains below 5 % for up to 50 vol% olivine, and it only exceeds 0.1 for olivine concentrations larger than 80-90 vol% (Fig. 5b and Table 2).

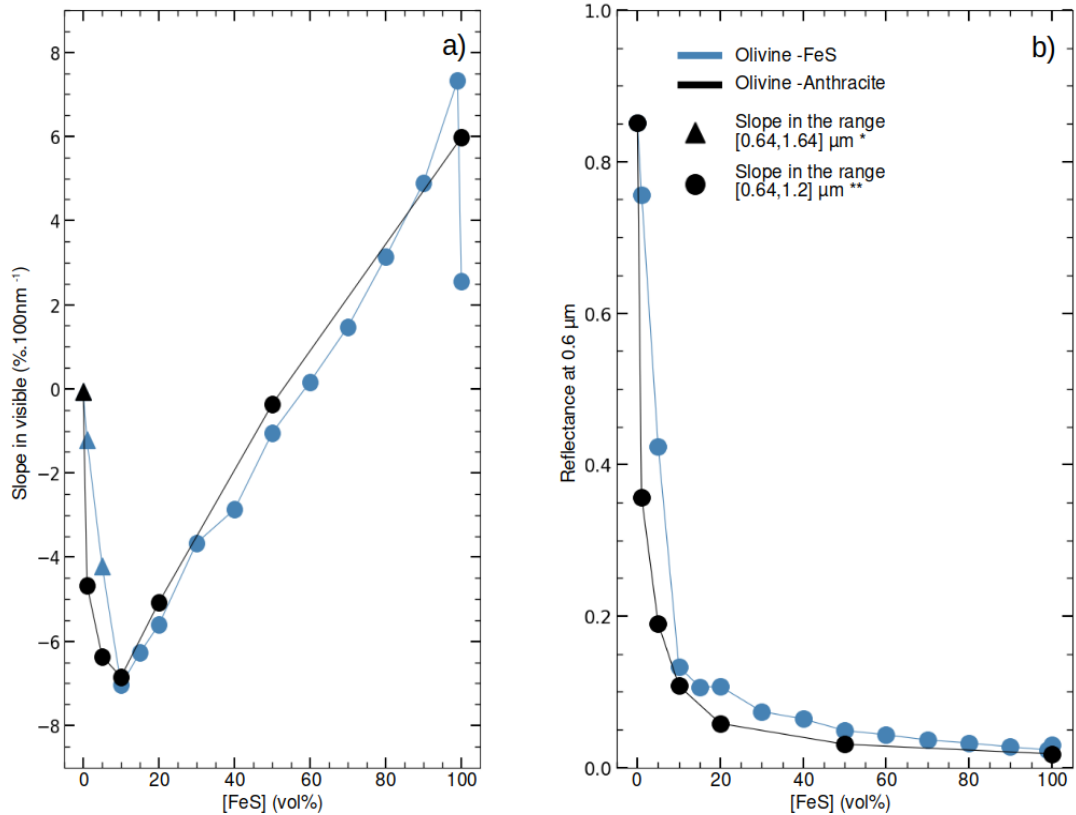

*Figure 5: Evolution of the spectral slope in the Vis-NIR (**a**) and of the reflectance at 0.6 µm (**b**) with increasing volume fraction of opaque grains in the samples. The slopes in the wavelength range ([0.64;1.64] µm) decrease rapidly with the volume fraction of olivine in the mixture for both types of mixture (FeS and anthracite). They reach minimal values of -7.0 and -6.8 %.(100.nm)$^{-1}$ for concentrations of 10 vol% of FeS and anthracite respectively. Then, for larger volume fractions of opaques, the slope increases progressively until it reaches the value of the respective pure opaque material spectrum. The reflectance is decreasing very fast with the amount of opaque grains in the mixtures, until they occupy ~10 vol% of the sample. It then decreases very progressively.*

**3.2 Grain size-effect on olivine MIR spectra and effect of dilution in KBr**

Figure 6 displays reflectance spectra in the mid-infrared for a suite of olivine samples with decreasing grain sizes, from hundreds of micrometres to sub-micrometre in diameter. Spectra first display an increase of the continuum reflectance in the inter-bands wavelength ranges, around 3.9 and 6.4 µm, as the grain size decreases This is due to an increase of the volume density of scatterers at the scale of the light penetration depth with decreasing grain size, leading to more scattering of incident photons relative to absorption. In the case of the hyperfine grains, the reflectance is overall much lower in the mid-infrared range. While reflectance maxima are observed around 9.5, 10.5, 16, 19 and 23 µm for grain sizes > 25 µm (where the strong stretching and bending modes of $SiO_4$ are located), the spectra obtained on olivine grains with hyperfine sizes show an overall low reflectance (below 5 % from 6 to 25 µm). One can observe that none of these spectra displays spectral signature in the 10-µm region similar to what is observed on small bodies (Vernazza and Beck, 2017).

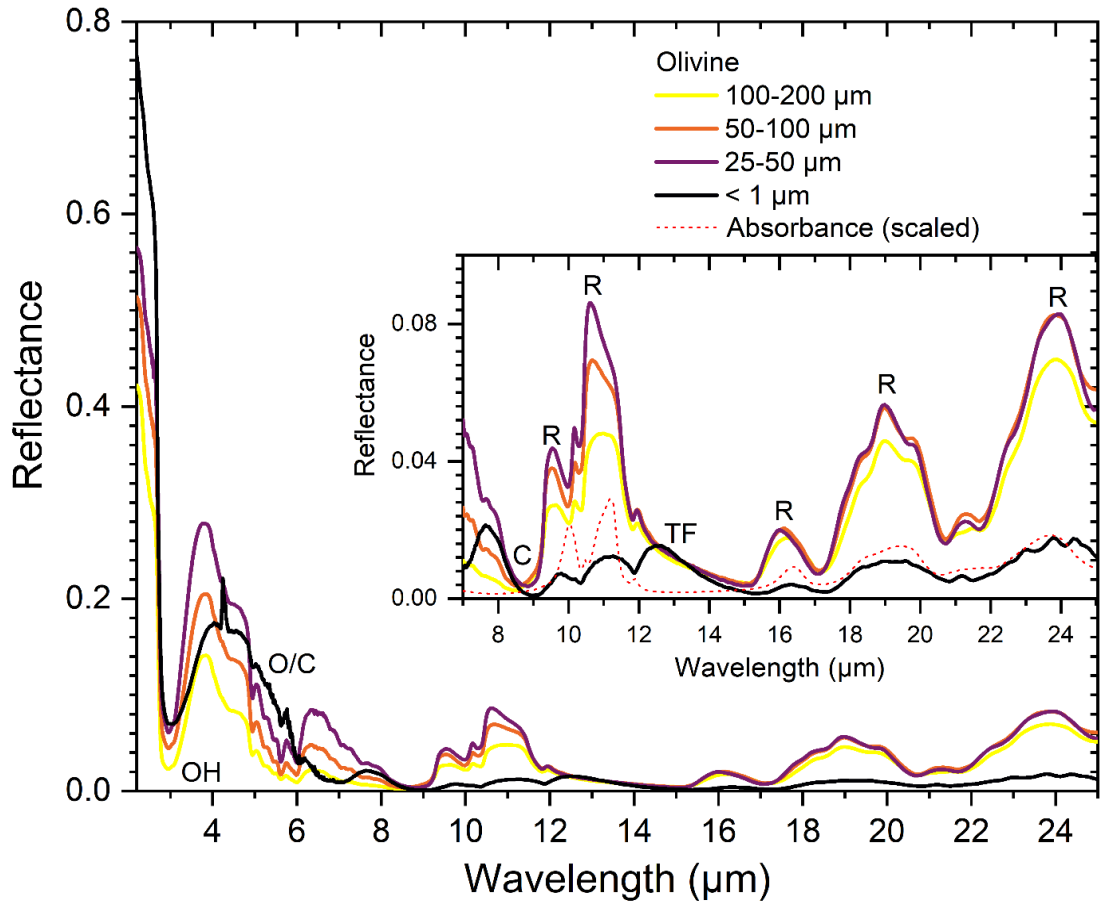

*Figure 6: Mid-infrared reflectance of olivine with decreasing grain size. The letters indicate the positions of different spectral features. OH: water and hydroxyl absorption bands; O/C: olivine overtone/combination absorption bands; C: Christiansen feature; TF: transparency feature; R: reststrahlen features. An absorbance spectrum of olivine, obtained by measuring the transmittance spectrum of a KBr pellet containing olivine powder, is shown for comparison. The reststrahlen features (R) tend to first increase with decreasing grain size, but for grains smaller than 1 µm, their reflectance level drops and the sample becomes very dark in the mid-infrared range. The small peak at 4.25 µm is related to gaseous $CO_2$.*

However, when the olivine is diluted in KBr powder as shown in Figure 7, we observe a spectral feature close to 10 µm in emissivity or "1-Reflectance" spectra. This feature consists in a double peak at 10 and 11 µm together with a smaller one around 12-µm. One can observe that this latter feature is the sharpest for the lowest concentration of olivine in the KBr matrix, and that the contrast of the feature with respect to the continuum decreases when the concentration of olivine increases. This can be interpreted by the fact that when mixed with KBr, the olivine sub-µm grains are de-agglomerated and dispersed in the sample, and more efficiently for lower concentration of olivine, so that they can absorb the light individually (and not collectively, as larger grains). The KBr crystals are non-absorbing and just scatter the light, inducing higher reflectance in the inter-bands wavelength ranges (transparency windows) and higher contrast. The spectral effects of mixing silicates with salts was studied in details by Izawa et al. (2021). In emissivity, the contrast increases with decreasing concentration of olivine because in the inter-bands, the thermal flux emitted from deeper layers is less efficiently transported to the surface. Therefore, a 10-µm plateau can be produced in "1-Reflectance" or emissivity spectra if a material "brighter" in the MIR and less thermally conductive is mixed with the hyperfine olivine.

Additionally, these measurements show that, in all three cases, the contrasts of the 10-µm plateau in emissivity spectra is lower than those calculated from reflectance spectra using the Kirchhoff's law. A likely explanation is that inverting biconical reflectance does not actually satisfy the conditions required for Kirchoff's law (Hapke, 2012), and does not give proper contrasts (Salisbury et al., 1991). Another possibility would be the presence of a thermal gradient in the sample, with increased radiance from hotter deeper layers reaching the surface in the inter-bands wavelength ranges (transparency windows), although this effect should be stronger at ~4-8 µm (more transparent) than at 13 µm, which is not always the case as seen in Figure 7.

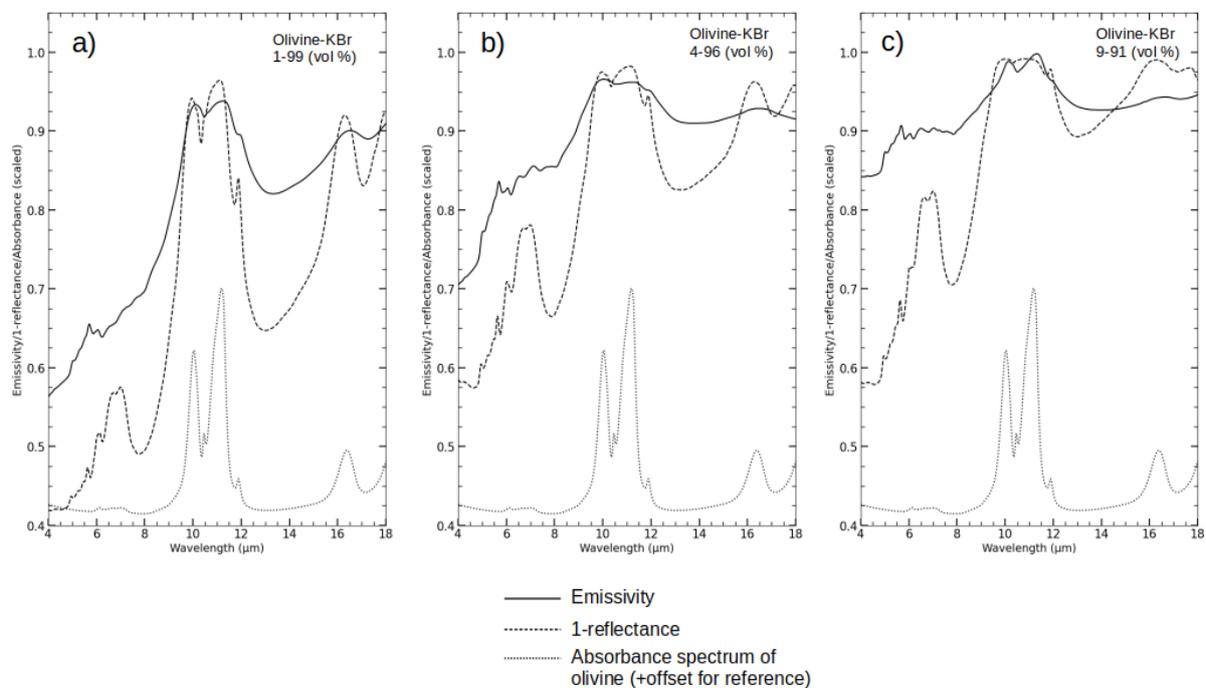

*Figure 7: Evolution of the emissivity contrast with olivine concentration in KBr measured directly in emissivity and in reflectance. The emissivity measurements presented here were performed at T = 400°C. Reflectance measurements were converted to emissivity using Kirchhoff's thermal law. One can observe that the lowest the concentration of olivine in KBr, the highest the emissivity feature around 10 µm. Moreover,*

*the contrast of the 10-µm band is lower in direct emissivity measurements than calculated from Kirchhoff's law, probably due to the influence of thermal gradients in the samples measured in emissivity.*

**3.3 Mid-infrared spectra of mixtures of hyperfine opaques and silicates**

Fe-rich opaques (Fe, FeS, Fe-Ni alloys) are absorbing in the VIS, but their reflectance spectra are significantly brighter than olivine in the MIR, probably due to their lower MIR absorption index k (e.g., Querry 1985 and Sato 1984, for Fe and $FeS_2$ respectively) (Fig. 2). They could thus be considered as brightening agents in the MIR. In Figure 8 and Figure 9 we present the emissivity spectra calculated with Kirchhoff law for the same mixtures of olivine and opaques studied in the VIS-NIR spectral range.

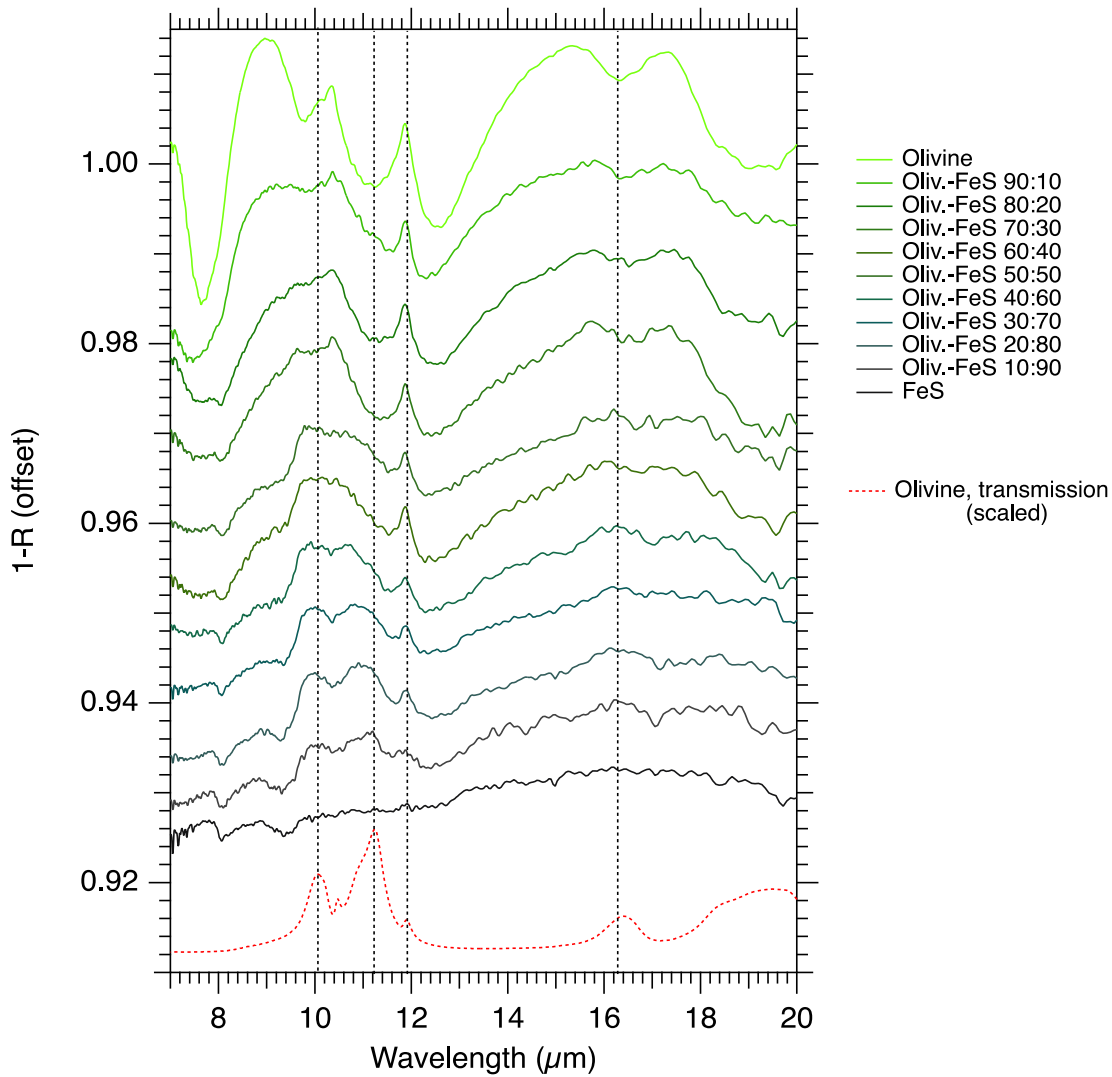

*Figure 8: Evolution of the MIR spectra of the olivine-iron sulfide mixtures. With the decreasing volume of olivine in the surface, we observe more pronounced silicate features with 2 components plateau between 9 and 12 µm. The original reflectance spectra are shown in Supplementary Figure 1.*

A relatively similar behaviour is observed for the two suites of mixtures (Figure 8 and Figure 9). While the spectra of the endmembers are relatively flat, the spectra of mixtures tend to show the presence of positive features in the 9 to 12 µm spectral range. Of interest is the appearance of the triplet around 10, 11 and 12 µm, where IR absorption bands are present in infrared transmission spectra of olivine. These observations reveal that mixing sub-µm grains of olivine and opaques (opaques in the visible spectral range) is a mechanism to produce an emissivity feature.

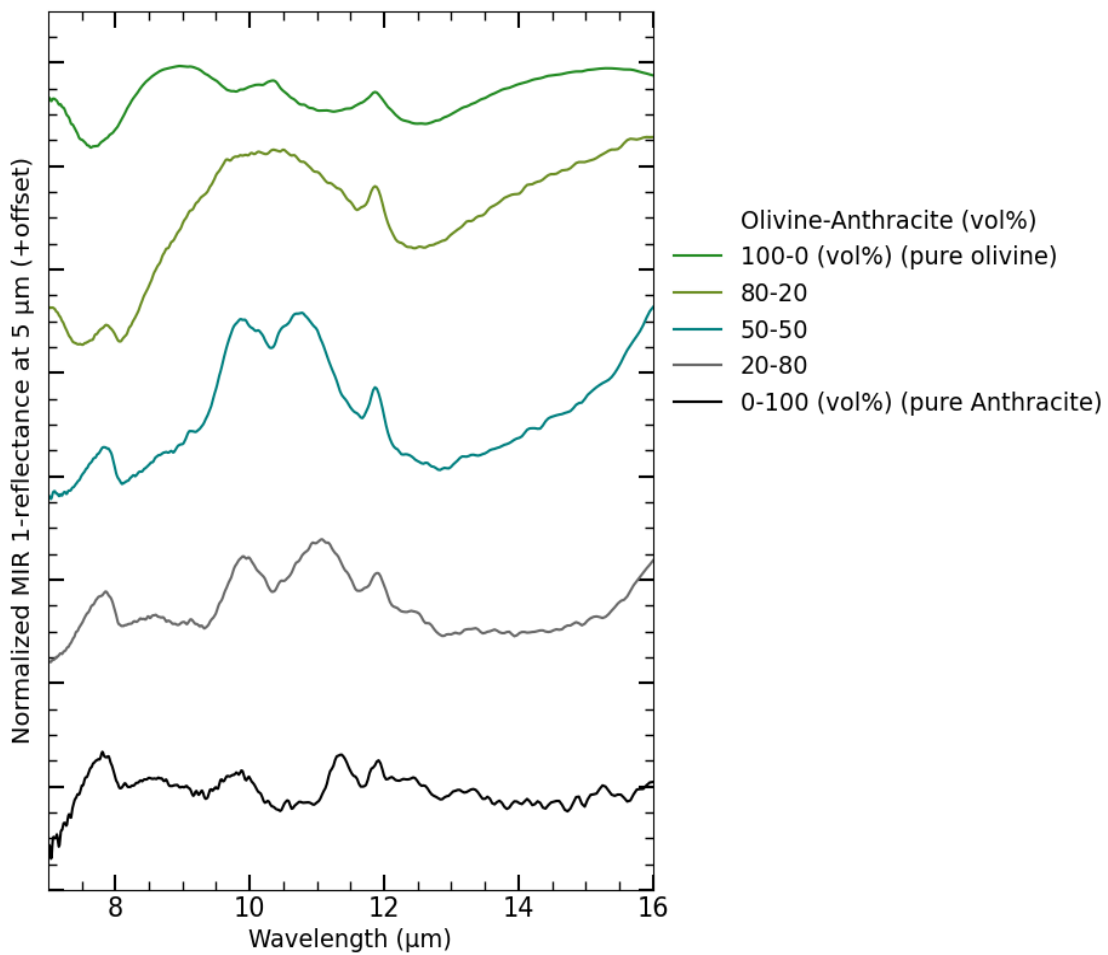

*Figure 9: Evolution of the MIR spectral shape of the mixture olivine-anthracite with the increasing volume fraction of opaque. With the decreasing volume of olivine in the surface, we observe more pronounced features of silicates with 2 components plateau between 9 and 12 µm. The original reflectance spectra are shown in Supplementary Figure 2.*

### 3.3 Linear polarization of visible light

Polarimetric phase curves were obtained for a selection of six olivine-FeS hyperfine mixtures as well as for the pure endmembers at the University of Bern. These polarimetric phase curves reveal the typical shape observed for particulate samples (in the laboratory or on planetary surfaces, Belskaya et al., 2015), with a "negative branch" at low phase angle (<30°) and a "positive branch" at higher phase angle (Fig. 10a, 10b). The behaviour and magnitude of polarization in the positive branch can be related to the contribution of photons that were reflected once at the surface of grains (Hapke, 2012). This can explain that the magnitude of polarization at high phase angle (> 50°) seems to be correlated with the reflectance of the sample (Fig. 10c). The higher the reflectance, the more contribution from multiple scattering can be expected, which explains the lower polarization.

In the analysis of polarimetric phase curves, the inversion angle (the angle at which linear polarization is null) and the absolute value of the minimum (referred to as $|P_{min}|$) have been used to empirically characterize the polarization properties of a variety of materials compared to observations of Solar system surfaces (Dollfus et al., 1989). The two endmember materials, FeS and olivine, display the shallowest negative branches with $|P_{min}|$ values of 0.25 % and 0.75 % respectively (Fig 10b, 10d). On the other hand, the pure FeS presents a polarization ratio of 25.4 % at 70° phase, and the pure olivine is the least polarizing material (with a polarization ratio of a few percent only at 70° phase).

The results obtained on the mixtures reveal a contrasted behaviour between the negative and positive branch of polarization (Fig. 10a,b). Reflected light from the pure iron sulfide is strongly polarized at high phase angle (25.4 %), but the strongest polarization (here 29 %, always occurring at 70°, the highest phase angle we measured) is obtained for the mixture with a small fraction of olivine (10 vol% olivine) (Fig 10a and 10c). Interestingly, the mixtures containing 1 to 20 vol% olivine have both stronger 530-nm polarization at 70° (Figure 10a) and redder Vis-NIR reflectance spectral slope (Figure 5a) than the pure FeS. This could be explained by the fact that olivine grains are more diffusing than FeS gains, so isolated sub-µm olivine grains at the surface of the sample constitute optical pits, diffusing sub-µm light onto more FeS-absorbing interfaces than when they are absent, decreasing the reflectance and increasing the positive polarization at sub-µm wavelengths. Then, with increasing olivine fraction, the maximum value of polarization seems to decrease monotonically with the fraction of FeS in the mixture, and is correlated with the visible reflectance.

The negative branch shows a highly non-linear behaviour with the opaque fraction in the mixture (Fig. 10b). The value of |$P_{min}$| is the strongest for the lowest fraction of FeS ($P_{min}$ = -2.5 % for 10 vol.% of FeS) and then progressively decreases (in absolute value) toward the value for the pure FeS (Fig. 10c). In the case of the inversion angle (Fig. 10e), it decreases from around 25° for pure FeS to about 21° for FeS:Olivine 90:10 vol%, then increases to its highest value for FeS:Olivine 80:20 vol%, and then decreases progressively toward the value for pure olivine (about 15°). These non-linear behaviours were previously observed for several types of mixtures by Shkuratov (1987), Shkuratov et al. (1992), and Spadaccia et al. (2022).

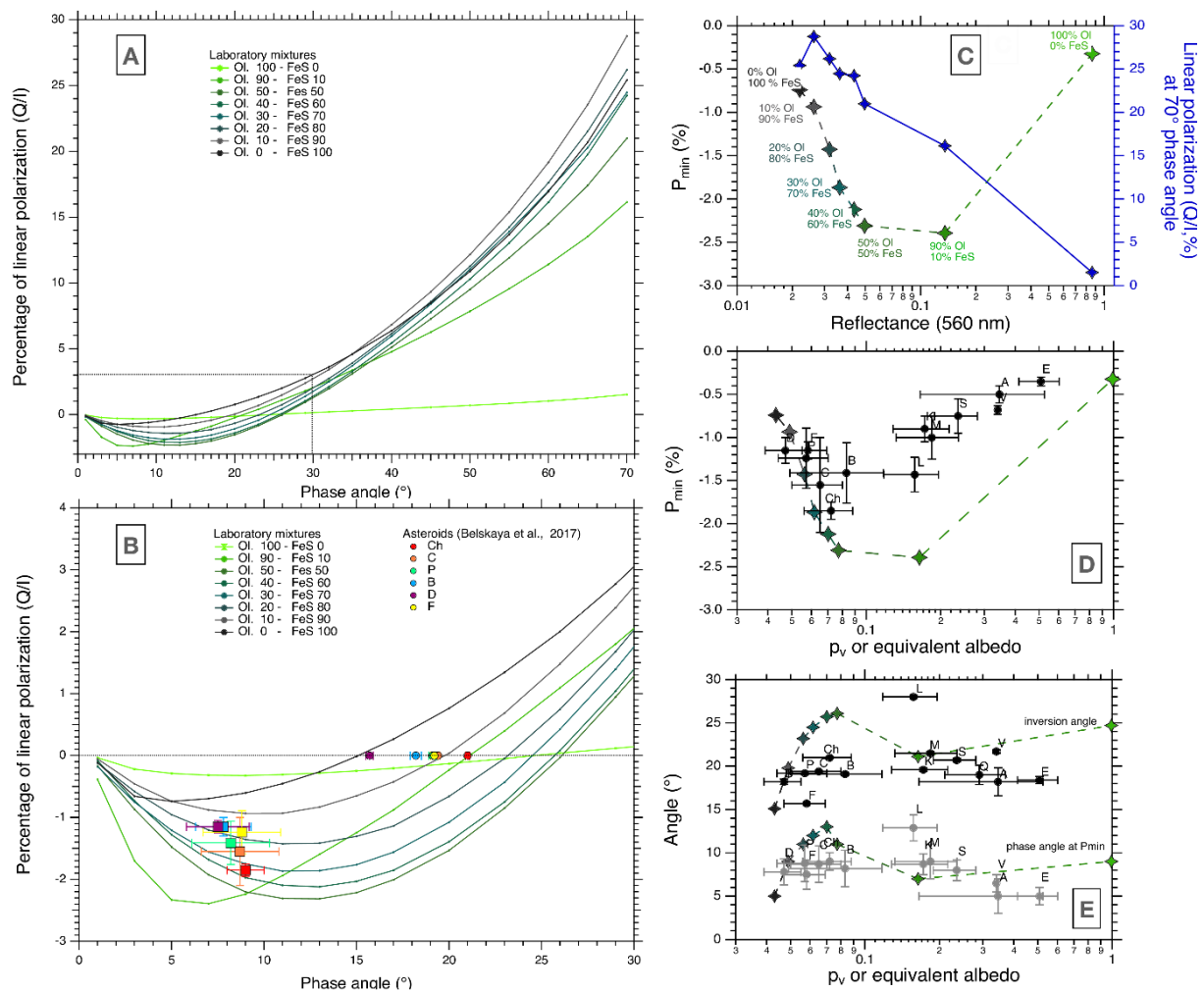

*Figure 10: Polarimetric phase curves obtained for the olivine-FeS mixtures (A). The bottom left panel (B) focuses on the negative branch, and values derived for P, D, B, C and Ch asteroids are taken from Belskaya et al. (2017). The three right panels are the calculated parameters for the polarimetric phase curves, also compared to asteroid data from Belskaya et al. (2017). The equivalent albedo was calculated using the law derived by Beck et al., (2021).*

# 4. Discussion

## 4.1 Rayleigh-like scattering as an origin of VIS spectral bluing

As described in the previous section and presented in Figure 3 and Figure 4, the spectra of mixtures of hyperfine grains of olivine and opaque material change with their volume proportions. When only few percent of opaques are present, the reflectance strongly decreases, the $Fe^{2+}$ absorption band at 1-µm of olivine disappears, and the VIS spectral slope is bluing in a relative sense (by "bluing", we mean that the relative spectral slope is getting more negative while the reflectance is decreasing at all wavelengths).

The influence of very fine opaques on reflectance spectra has been particularly studied in the framework of space weathering (Hapke, 2001; Noble et al., 2007; Pieters and Noble, 2016). The incorporation of < 25 nm opaques within larger translucent grains (intra-mixture) induces a fast decrease of reflectance, associated to a reddening of the spectra (Hapke, 2001, Noble et al., 2007). The incorporation of 25-200 nm opaques induces a reddening or a bluing depending on the concentration (Noble et al., 2007). Our mixtures with ~300 nm hyperfine opaque grains (inter-mixture) present the same evolution of the reflectance, with an efficient darkening of the samples, but on the contrary only a bluing of the spectra. The spectral slope variation trend reported in Figure 3 and Figure 4 for two different types of opaques are similar, indicating that this bluing phenomenon is independent from the composition of the opaque material, so a physical cause is to be searched for. Note that the two opaque grains used here are of similar sizes (cf. Table 1, 0.30 µm for FeS and 0.33 µm for anthracite).

To explain the observation of a blue peak in the spectra of Saturn's moons, (Clark et al., 2008), measured reflectance spectra of mixture of ice with 0.2 µm diameter sized particles of carbon black. Clark et al., 2008 showed that these mixtures display a blue peak around 0.5 µm increasing in intensity with the amount of the contaminant, then decreasing after a threshold of concentration. This study also indicated that this peak was not dependent on the contaminant spectrum, but that it was related to light scattering in a Rayleigh regime.

Following Clark et al. (2008) we propose that in our sample the hyperfine grains of opaques, well dispersed in a bright transparent matrix of silicate grains, can scatter light following a Rayleigh-like regime, leading to a bluing of the spectra; lower wavelengths are more efficiently scattered than higher wavelengths. In order to test this interpretation, the theoretical approach of Brown (2014) was used, where the author calculated the parameter space in term of optical constant and size parameter where spectral bluing should occur. We reproduced the map of maximum variation of the single scattering albedo $\varpi$ as a function of the size parameter $X$ and the imaginary part of the optical index $k$, to assess if our grains are situated in the region of expected bluing in the $X$-$k$ space. To do so, optical indexes of olivine, anthracite and iron sulfide are needed. For the two first materials, we respectively used the indexes from Zeidler et al. (2011) and Blokh and Burak (1972). For iron sulfide, as there is no optical indexes available in the Vis-NIR for either pyrrhotite or troilite, we decided to approximate it by the optical indexes of metallic iron, using the indexes from Querry (1985).

With all the optical indexes and the mean grain size of our hyperfine powders, we could compute the first derivative to the Rayleigh approximation of the single scattering albedo as established in Brown (2014):

$$\frac{d\varpi}{dX} = X^2$$

The results of this computation are shown in Figure 11. For each material, we computed the size parameter $X$ and the first derivative of the single scattering albedo at 0.7 µm for each material. On Figure 11, we observe that olivine is outside of the bluing region, but that the anthracite and the iron sulfide are very close to it in the $X$-$k$ space. If we consider their size distribution (displayed as error bars on Fig. 11), it appears that both types of opaque grains are crossing the region of spectral bluing. This confirms that the strong blue slope appearing in our mixtures takes its origin in the light scattered by absorbing and optically isolated hyperfine grains.

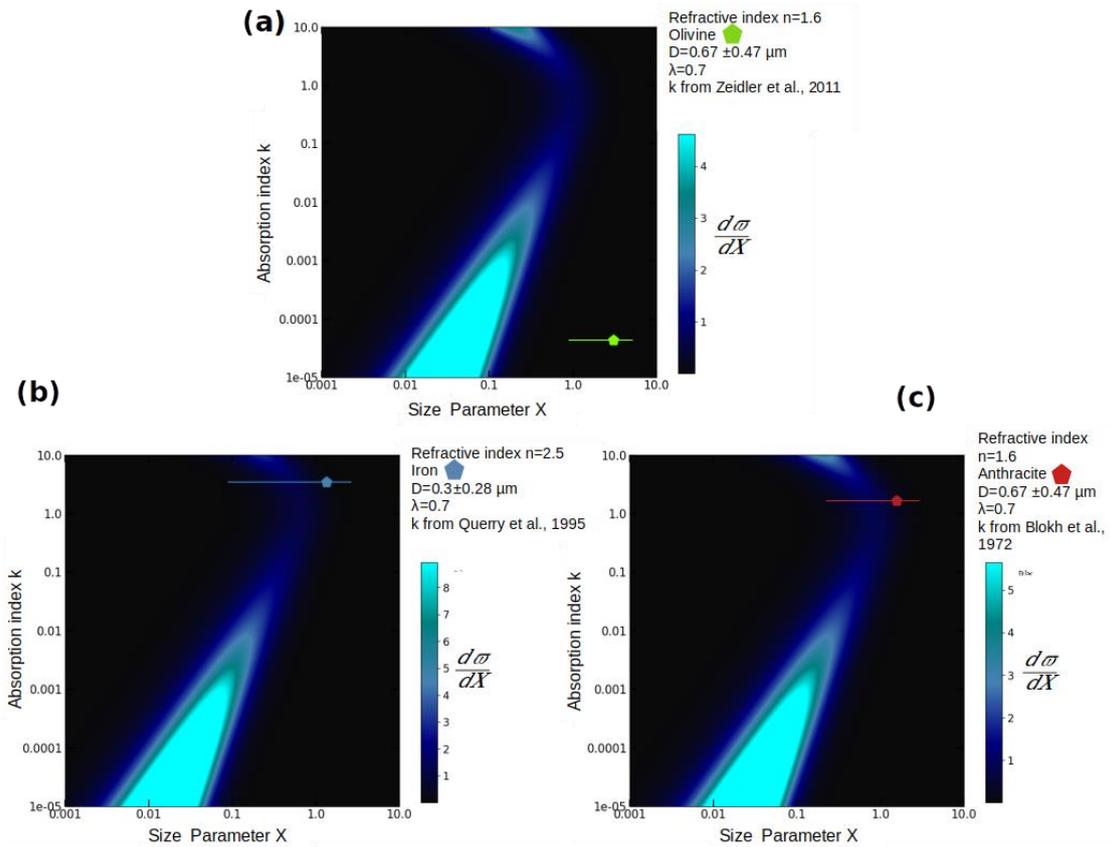

*Figure 11: Evolution of the spectral Rayleigh bluing region in the X-k space following the work of Brown (2014) for optical indexes of olivine (a), iron (b) and anthracite (c). The figure indicates that the spectral bluing is restricted to a well-delimited curved region in the X-k space (higher values of $\frac{d\varpi}{dX}$ means more bluing). The mean grain size and the range covered by the size distribution are indicated for each sample. Some of the grains present in the powders of the two opaque materials are crossing the bluing region because of the size distribution (b, c), indicating these grains are interacting with light following the Rayleigh regime. On the contrary, olivine grains are outside of the bluing region.*

To assess whether olivine grains act as a non-absorbing matrix in the mixture, we computed the ratio of the absorption $Q_a$ and scattering $Q_s$ cross sections in the same *X-k* space for the three materials (Figure 12). To this aim, we used Mie theory and a Python implementation coded by (Sumlin et al., 2018). On Figure 12 we position the grain size distributions of olivine, iron sulfide and anthracite in the *X-k* space indicating regions where grains are more absorbing than scattering light. Olivine grains are situated in a region where light is more scattered than absorbed, whereas the grains of anthracite and iron sulfide are in an absorbing region. This suggests that in the visible range, light is absorbed following the Rayleigh regime by the absorbing grains of opaques, and that the olivine acts as a matrix of transparent materials.

The presence of the olivine grains is nonetheless essential for the bluing phenomenon to be observed, because they allow the separation of the individual hyperfine opaque grains from each other (see Fig. 2). In the pure opaque powder, the optical scatterers' size may be larger than the physical size of grains, and may correspond to the size of agglomerates rather than individual grains (this was observed in Sultana et al. (2021) for bright materials). The mixture with olivine grains allow the de-agglomeration of the opaque grains, which can then scatter the light individually and produce the spectral bluing.

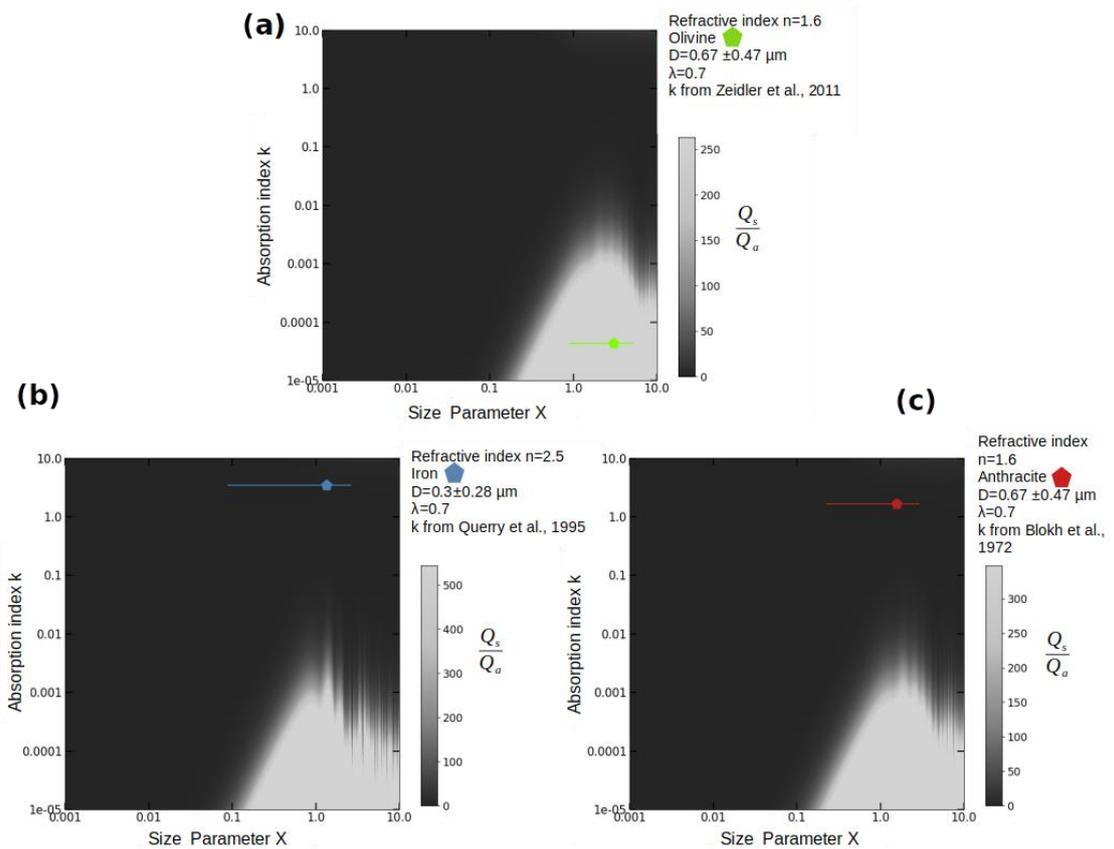

*Figure 12: Ratio between Mie scattering and absorption efficiency in the X-k space at a wavelength of 0.7 µm. These plots indicate the region of this X-k space where light scattering is predominant during interaction of light with the grains. We placed our grains in this X-k space. It appears that anthracite and iron sulfide are absorbers whereas the olivine is mostly scattering light.*

The fast disappearance of the olivine $Fe^{2+}$ absorption band at 1 µm with increasing proportion of opaques can be explained by the reduction of the optical path lengths through the olivine grains, induced by the presence of the absorbing opaque grains. Even for the small grain sizes investigated here, the single scattering albedo of the opaque grains is low, and the encounter between a photon and an opaque grain will lead to its extinction.

**4.2 Polarimetry of mixtures of hyperfine grains**

The properties of the polarimetric phase curves (inversion angle $α_{inv}$, $P_{min}$, slope at inversion $h$) for powders of minerals and meteorites of various grain sizes and albedo have been studied historically by Dollfus, Shkuratov and colleagues. Based on a $P_{min}−α_{inv}$ diagram, Dollfus et al. (1989) proposed that most asteroids are covered by a coarser regolith (~30-300 µm) compared to the lunar surface. Based on the same diagram, the low inversion angles recently measured for F and D-type asteroids (Belskaya et al., 2005; Bagnulo et al., 2016) would suggest that their surfaces are covered with "bare rocks". However, their peculiar polarimetric properties may also be explained by their microstructure. Indeed, laboratory measurements performed by Shkuratov (1987) have shown that mixtures of powders having grains ≤ 1 µm and very contrasted albedos exhibit an amplified negative branch of polarization compared to powders of the single components. Other measurements reported in Belskaya et al., 2005 demonstrate that adding small amounts of a bright powder (e.g., MgO) to a dark powder (e.g., pure carbon soot) noticeably increases $P_{min}$ and $α_{inv}$, while producing a very faint albedo increase. Interestingly, the polarimetric slope h is recognised to be a good indicator of the albedo, except for very dark surfaces. Laboratory measurements showed that the empirical relationship between h and albedo began to "saturate" for very dark surfaces (albedos of about 0.06 or lower) made of mixtures of very small grains (≤ 1 µm), principally dark ones mixed with a small portion of bright ones. Zellner et al. (1977) first reported that h remains approximately constant when the albedo of mixtures decreases. For darker mixtures, Shkuratov et al. (1992) reported that as the albedo decreases, h and $P_{min}$ reach a maximum before decreasing and $α_{inv}$ also decreases. This "saturation" effect is apparent in the observations of several F-type and D-type asteroids (Belskaya et al., 2005; Cellino et al., 2015; Bagnulo et al., 2016), *suggesting* that they have a peculiar surface structure at the (sub)-micrometric scale compared to other asteroids.

Our experiments with mixtures using small bodies analogue constituents are in agreement with earlier findings, that the intimate mixture of materials with strongly contrasted absorption properties induces an amplified polarization in the negative branch (Fig. 10b). We also observe the saturation effect described by Shkuratov et al. (1992). There is a maximum of |$P_{min}$| for olivine contents in the 90-50 % range, followed by a progressive decrease of |$P_{min}$| for higher FeS content (Fig. 10c). Remarkably, the introduction of a small fraction of FeS in olivine changes drastically the negative branch, with a high value of |$P_{min}$| and a lowering of the inversion angle. The reflectance at 0.7 µm is decreased from around 0.85 to 0.13 from pure

olivine to the 90:10 % mixture, showing that the FeS fraction has a strong influence on the radiative transfer, which explains the major changes in polarimetric properties. With further increase in the FeS fraction (50-50 %), the inversion angle increases and then we reach the "saturation region" where |$P_{min}$| and $\alpha_{inv}$ decrease with increasing FeS content.

Inspection of the polarimetric parameters for small bodies of different types seems to reveal a similar saturation behaviour. As the albedo decreases from E-type toward L-types, the |$P_{min}$| values increase progressively (Belskaya et al., 2017). The highest value of |$P_{min}$| are encountered for the Ch spectral types, which experienced significant aqueous alteration, and are likely chondrules-bearing given their spectral connection to CM chondrites (Vernazza et al., 2016). Then |$P_{min}$| values seem to decrease when considering C, P, D and F-type asteroids that have slightly lower albedos. Interestingly, these spectral types land nicely on the trend defined by our mixtures in the $p_v$ vs |$P_{min}$| diagram (Fig. 10d), in the locations defined by FeS-rich mixtures (80 or 90 % FeS). In the case of the P and D-type, the curves we obtained for the 90 % FeS mixture also seem to match the inversion angle.

While the FeS content is high in the 90 % FeS mixture, and Fe/Si and S/Si are in excess of a solar-like composition, we propose that the combination of texture (hyperfine grains) and the cohabitation of weakly and strongly absorbing constituents explains the polarimetric properties of P and D-type small bodies. These constituents may be various types of silicates, opaque minerals (iron sulfides, Fe-Ni alloys, oxides etc.) and organic matter. We can also remark that pure iron sulfides show polarimetric properties that differ from typical values for expected-to-be-metallic M-type asteroids (Belskaya et al., 2022), but also show a quite different visible reflectance. This suggests that the surfaces of metallic asteroids are not covered only by metallic fines, but by a coarser-grained regolith, possibly silicates-bearing.

**4.3 B-type asteroids and Rayleigh-like scattering**

Objects belonging to the B-type asteroid population are by definition objects with a negative spectral slope. A now famous example of B-type object is the near Earth asteroid (101995) Bennu (the target of the OSIRIS-Rex mission) that displays a blue slope in the Vis-NIR (Clark et al., 2011; Hamilton et al., 2019a) and lacks any absorption band until 2.7 µm where an absorption band related to hydrated minerals can be seen (Hamilton et al., 2019a). Other notorious examples of B-types are the asteroids (2) Pallas and (3200) Phaethon (Kareta et al., 2018a). Rayleigh-like scattering is one of the possibilities that have been formulated to explain the peculiar B-type spectra (Beck and Poch, 2021), following experiments designed to explain the spectral bluing of selected crater ejecta on 1-Ceres (Schröder et al., 2021).

Besides being spectrally blue, the spectra of B-type asteroids present a convex shape, reminiscent of the Rayleigh-like scattering observed for some of our samples. In Figure 13a we plot the normalized spectra of these three B-type asteroids against the most spectrally similar mixtures of olivine and iron sulfide that we prepared. Pallas and Phaethon are spectrally bluer than Bennu, which would be consistent with a higher volume fraction of opaque minerals on Bennu (but other explanations are possible: this could also be due to different constituents having different optical indexes, or to different rock porosities).

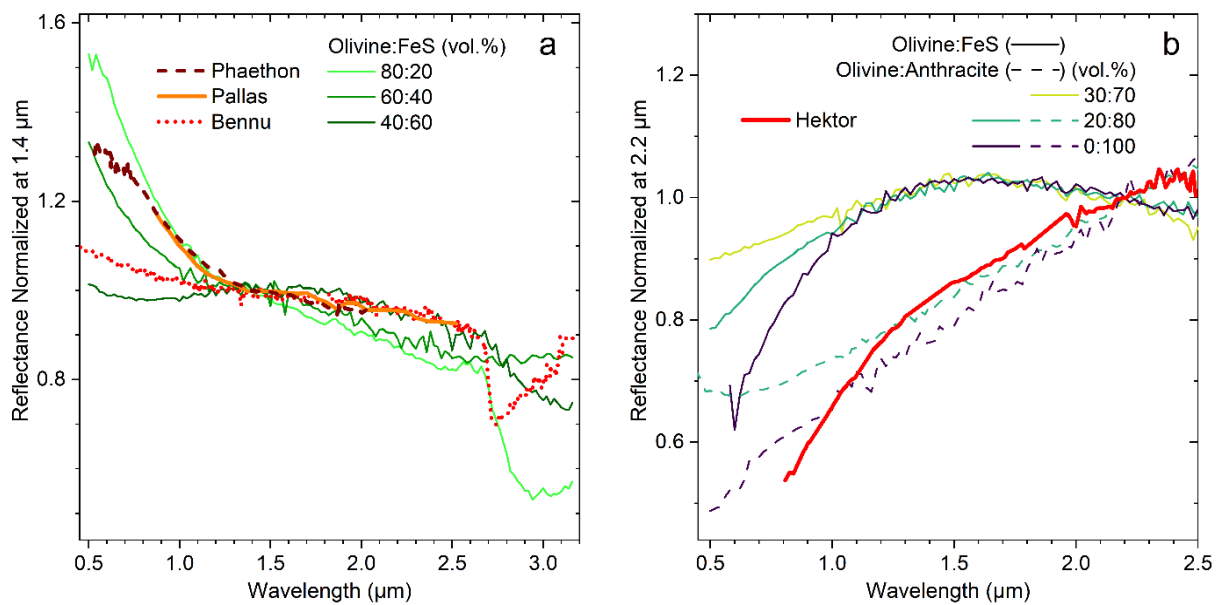

*Figure 13: Comparison of normalized Vis-NIR reflectance spectra of asteroids with some of mixtures of sub-µm grains. **(a)** Vis-NIR spectra of three B-type asteroids (101995) Bennu (Hamilton et al., 2019a), (2) Pallas (DeMeo et al., 2009) and (3200) Phaethon (Kareta et al., 2018a) and the most similar ones in terms of spectral shape measured on Olivine-FeS sub-µm mixtures, all normalized at 1.4 µm. Pallas and Phaethon are spectrally bluer than Bennu, which would be consistent with a lower volume fraction of opaque materials on their surfaces (but other explanations, for example differences of constituents or porosity, are also possible). **(b)** Vis-NIR spectra of the D-type Jupiter Trojan (624) Hektor (Emery and Brown, 2003) and the most similar ones in terms of spectral shape measured on Olivine-FeS and Olivine-Anthracite sub-µm mixtures, all normalized at 2.2 µm. The spectral shape varies strongly depending on the nature of the opaque material, and only the mixtures containing very high proportions of opaques have spectral slopes closest to D-type asteroids. The spectral slope of D-types may be explained by the presence of other constituents, possibly having different size distributions, and/or different micro-porosity, and/or by space weathering of their surface.*

However, a proportion of about 40 to 60 vol% of opaque minerals on Bennu is significantly higher that what is found for proposed spectral analogues of Bennu. These analogues includes CM and CR Chondrites (Hamilton et al., 2022), whose FeS fraction is typically of the order of several to 10 vol% (Howard et al., 2011). Nevertheless, it is well known that the reflectance spectra of intimate binary mixtures are extremely dependent on the relative grain size of the two constituents (see for instance Pommerol and Schmidt, 2008). Mixtures of hyperfine opaques with larger grained silicates may produce a spectral bluing, together with a darkening, for a much lower fraction of opaques. We should also remark that while spectral analogues of Bennu may be found in databases, spectra obtained on textural analogues are scarce if not absent in the literature. Based on orbital observation, the surface of Bennu resembles a rubble-pile, that seems to be covered by angular rocks rather than fine grained-regolith. However, mid-infrared observations suggest that these rocks are substantially porous (Rozitis et al., 2020). As a consequence, meteorites, that need some strength to survive atmospheric entry, do not seem to provide good textural analogue. Our mixtures are unconsolidated aggregates of hyperfine grains, but they show that the presence of dispersed hyperfine grained opaques in these porous rocks could provide an explanation for the blue slope of Bennu. We note that consolidation of particulate media can induce darkening (Schröder et al., 2017), and possibly other spectral changes, which would need to be studied further.

**4.4 Hyperfine grains and MIR emissivity features**

In the MIR spectral range, silicates present strong optical absorptions related to Si-O vibrations (stretching around 10 µm and bending at longer wavelength). As a consequence, silicate absorptions bands of powders with "large" grain size (> 20 µm) observed in this spectral range are often an interplay of surface and volume scattering, as is usually the case for strong absorption bands (Hapke, 2012). Also, grain size will not only change the intensity of the spectral features (such as restrahlen bands, transparency bands, and the Christensen feature), but also their position (Salisbury and Wald, 1991; Mustard and Hays, 1997).

In the case when grain size is decreased below the wavelength, the scattering efficiencies of grains decreases, as well as their single scattering albedo. Consequently, photons will penetrate deeper in the samples, and have an increase likelihood to be absorbed before exiting the sample toward the observer. As studied experimentally in Mustard and Hays (1997), and confirmed here for even smaller grains, decreasing grain size below the wavelength leads to an overall flat and very low reflectance of olivine in the mid-infrared spectral range. In the case of our hyperfine olivine powder, reflectance is below 3 % above 6 µm.

In the case of the hyperfine powders of "opaques" studied here, iron sulfide and anthracite, the behaviour should be virtually identical to that of olivine. If the grains size is small enough, we might also expect a Rayleigh-like behaviour for individual grains, leading to a very low reflectance of the sample. However, the reflectance of our pure opaques is not as low as the hyperfine olivine. This is possibly explained by the fact that the critical grain diameter, the diameter for which a powder shows a strong decrease of reflectance (Mustard and Hays, 1997), depends on the refractive index of the material. Another important point is that we are studying particulate samples, and that particles may behave collectively. A photon of 10-µm wavelength may sense a continuous medium (i.e. effective media theory) rather than a "cloud" of isolated hyperfine grains.

When hyperfine olivine is mixed with KBr, it was found here that the resulting reflectance spectra show pronounced spectral features of olivine, which would result in a strong 10-µm emissivity feature according to Kirchoff's law. This effect can be explained by the fact that the large KBr grains are dispersing and isolating the hyperfine grains of olivine and they scatter light, unlike larger olivine grains, enabling photons to escape the sample. Because some of these escaping photons will have interacted with olivine grains, the olivine signature is imprinted on the measured signal. We should then remark that using KBr powder to simulate porosity is not a valid approach, since while not absorbing, it still strongly influences radiative transfer through scattering. In order to simulate porosity with KBr, one needs to prepare the sample as pressed pellet in order to minimize scattering by the salt.

We propose that the effect of mixing olivine with sulfide and anthracite is somehow similar to that of mixing with KBr powder, but with a smaller magnitude. Moreover, the sub-micrometre-sized olivine grains are de-agglomerated and isolated from each other when mixed with opaque grains, as in a "cloud" of isolated hyperfine olivine grains. The opaque grains enable photons to escape the sample, and some of those will have interacted with olivine grains in the mixture. As a consequence, the obtained spectrum is reminiscent of an olivine absorption spectrum as seen in Figure 8.

### 4.5 On the origin of emissivity spectral features on primitive small bodies

Mid-infrared spectra of many objects of different spectral types (C, D- and P-types asteroids, comets) and from several dynamical reservoirs (Jupiter Trojans, Main Belt) exhibit 10-µm emissivity features. In Figure 14 we compare emissivity spectra from astronomical observations of several objects to "1-Reflectance" spectra of our mixtures of olivine and iron sulfide. According to the Kirchoff's law, "1-Reflectance" corresponds to the emissivity if the measured reflectance is the hemispherical reflectance (Hapke, 2012). Here, we have measured the biconal reflectance so band positions should be the same, but contrasts (absolute and relative) can be different compared to emissivity spectra (Salisbury et al., 1991). We can observe a fair match of the 10-µm feature between the observations and the mixtures containing between 10 to 40 vol% olivine. We can however notice that the emission feature is broader for the astronomical observations than for these mixtures. This difference is likely due to the presence of amorphous silicates (see for instance models of ejected dust from Tempel-1 by Gicquel et al., 2012) and crystalline pyroxene (present in anhydrous CP-IDPs, see Brunetto et al., 2011) on P/D-type asteroids and in cometary dust, which are absent in our samples.

The samples studied here have macro- and micro-porosities lower than 78%, more compact than "fairy castle" hyperporous (80-99%) ones, but still exhibiting a silicate signature at 10 µm as observed in MIR spectra of P/D-type asteroids and cometary dust tails. We propose that, in both cases, an optical isolation of olivine grains occurred, either by vacuum in the case of cometary dust tails, or by opaque grains and vacuum in the case of P/D-type asteroids. We should note that the strength of the emissivity feature we measured in the laboratory is at the percent level. As seen in Figure 14, this is of the order of what is observed for some D-type objects like (515) Athalia (~3-4 %, Licandro et al., 2012), but not as high as the strength of the emissivity feature observed for some Trojans (of the order of 10 %, Emery et al., 2006). The relative grains sizes of the opaque and silicate materials may have an impact on the intensity of the feature. In addition, while high surface porosity does not seem to be required to produce an emissivity feature, it may have an impact on its magnitude.

**4.6 On the origin of red spectral slopes on primitive small bodies**

The mixtures containing 10 to 40 vol% olivine are the ones exhibiting a 10-µm feature (Figure 14) and also a red (positive) Vis-NIR spectral slope (Figure 5a, Figure 13b). However, as for the 10-µm feature, the shape and magnitude of the Vis-NIR spectral slope are different from the ones of P/D-type asteroids (Figure 13b). Only mixtures with excessively high proportions of opaque grains (80-100 vol%) reproduce the magnitude of the spectral slope. These proportions are unrealistic (CP-IDPs contain ≤ 40 vol% Fe-Ni sulfides, Bradley, 2014), therefore other parameters control the Vis-NIR spectral slope of these bodies. These parameters may be all or some of the following: other types of spectrally red and opaque materials, space weathering reddening grain surfaces, higher micro-porosity, and different grain size distributions than the samples studied here.

In addition, the mixtures containing more realistic proportions of opaque grains do not show the 1-µm absorption band of olivine (Figure 3, Figure 4). Following these measurements, it is normal that the 1- and 2-µm absorption bands due to silicates are absent on P/D-type asteroids and comets spectra.

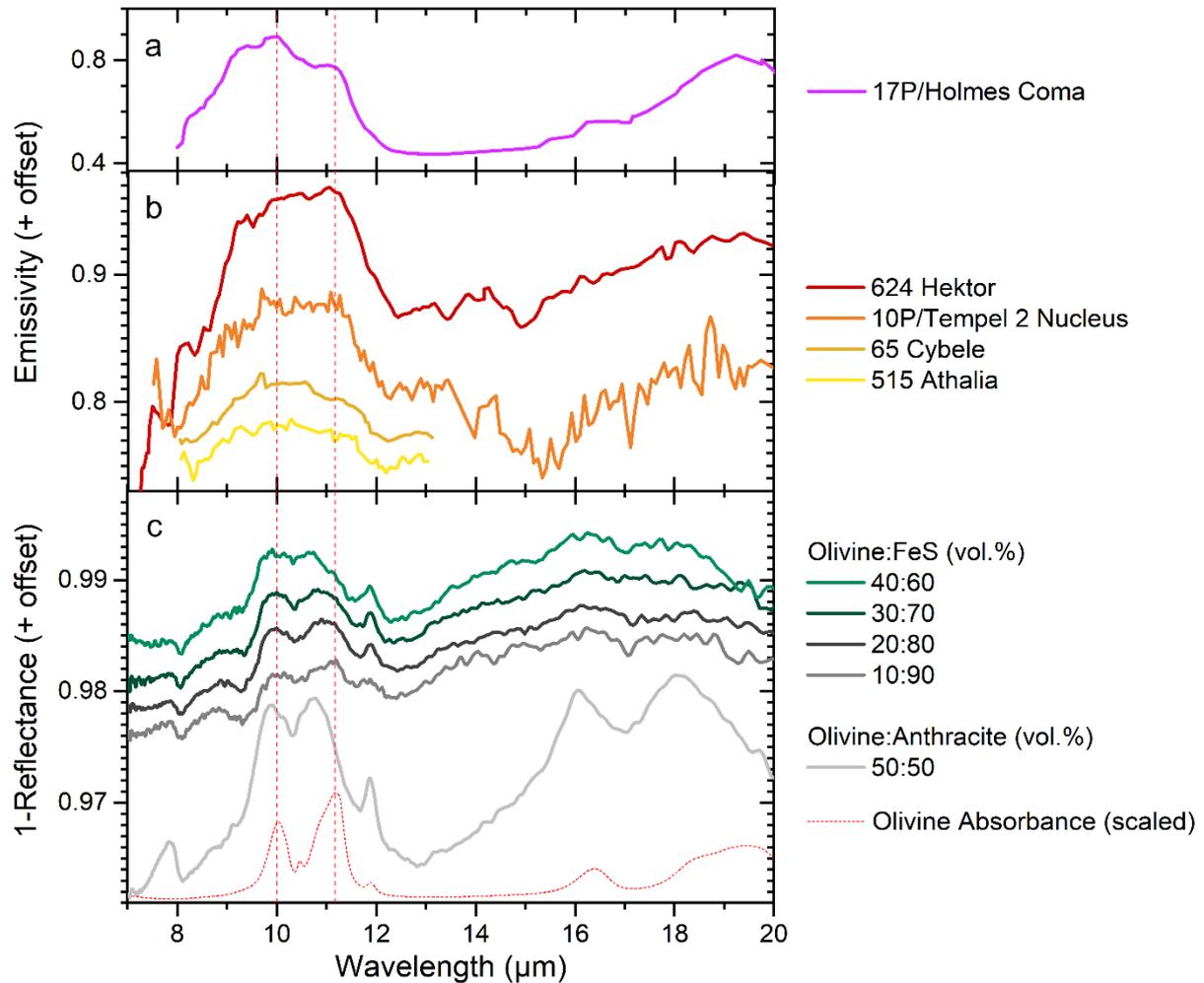

*Figure 14: Mid-infrared emission spectra of the coma of 17P/Holmes from Reach et al. (2010) (panel a), of surfaces of a comet (10P/Tempel 1, from Kelley et al. (2017)) and of C and D-type asteroids (the spectra are from Emery et al. (2006) for the D-type (624) Hektor, and Licandro et al. (2012) for the C-type (65) Cybele and (515) Athalia) (panel b), compared to "1-Reflectance" spectra of mixtures of sub-µm grains of olivine with iron sulfide or anthracite (panel c). The 10-µm emissivity plateau is prominent on the asteroids' spectra. A 10-µm plateau is also observed at a similar position on the spectra of the mixtures containing 10 to 50 vol% olivine. The vertical red dashed lines indicate maxima of absorption of olivine. We can notice that the plateau on the asteroids and comets spectra extends to lower wavelengths (from 8 to 12 µm) than on the measured spectra (from 9.5 to 12 µm). This is related to the presence of other types of silicates, widening the 10-µm feature on small bodies spectra. The contrast between the plateau and the continuum is at the percent level for the mixtures, and ranges from 1 to 10 % for small bodies. These differences may be due to variations of relative grain sizes of opaques and silicates, and/or of porosity.*

## 4.7 Summary: combination of Vis-NIR, MIR spectral and Vis polarimetric properties

With the Vis-NIR, MIR spectral and Vis polarimetric measurements in hand, we can try to retrieve information regarding surfaces compositions and textures of B, C and D type objects.

By comparing the measurements from Vis-NIR and MIR, we observe that the mixtures exhibiting blue-sloped spectra in the Vis-NIR also lack a 10-µm emissivity plateau in the MIR, whereas surfaces presenting reddish slopes in the Vis-NIR exhibit a 10-µm emissivity plateau in the MIR. In Figure 15, we compile optical spectra in the Vis-NIR and in the MIR of cometary nuclei and asteroids. Similarly to our analogues, we observe an anti-correlation between the presence of the plateau in the MIR and the blue slope in the Vis-NIR when conjugating the two spectral ranges. Bennu spectrum in the Vis-NIR is indeed characterized by a blue slope between 0.4 and 2 µm, but its MIR spectrum is lacking any emissivity plateau. On the contrary, the Trojan asteroids, spectrally red in the Vis-NIR, do present a highly contrasted plateau between 8 and 12 µm.

These experimental data reveal that optical separation of sub-µm grains is a major parameter controlling the optical properties of low-albedo small bodies. For the samples considered here, this optical separation is obtained by mixing at extreme proportions two sub-µm materials having contrasted optical indexes in the Vis-NIR and MIR. But such an optical separation may also be provided by the porosity of the medium. Surfaces made of sub-µm opaque grains optically isolated (by a matrix of brighter grains, and/or by porosity) tend to have a blue spectral slope in the Vis-NIR and lack a 10-µm plateau. As larger aggregates of opaque grains are formed (by increasing their proportion in mixture, and/or by decreasing the medium porosity) they do not scatter light in the Rayleigh regime anymore, but they absorb the light, and the spectra become more neutral to red. These opaque grains responsible for the absorption in the Vis-NIR are more reflective than silicates in the MIR, so that surfaces made of sub-µm silicate grains optically isolated (by these more reflective grains, and/or by porosity) tend to have redder Vis-NIR spectra and display a 10-µm plateau.

Polarimetric parameters can be further added to this comparison. In the case of D-type objects, the mixtures that are best spectral analogues in the Vis-NIR and MIR (Fig. 15) seem to have close polarimetric properties as well (Fig. 10). The fact that the combination of these three optical properties provides a good match to the observation of D-type asteroids builds a very strong case for the fact that heterogeneous aggregates of hyperfine grains having contrasted optical properties are essential for explaining the D-type asteroid optical properties.

In the case of B-type asteroids, the polarimetric properties of the best spectral analogues among our mixtures (olivine-FeS from 40:60 to 70:30 vol%, Fig. 15) display higher values of |$P_{min}$| and $α_{inv}$ than ground-based observations of these objects (Fig. 10). This may suggest that the bluing may not be related to Rayleigh-like scattering for B-type asteroids, or that another phase or process plays a role in their polarimetric phase dependence without significantly impacting their reflectance spectra. In the case of Phaethon, polarimetric curves up to high-phase angles have been obtained (Ito et al., 2018) and the positive branch is well constrained. The polarization degree is of about 20 % for a phase angle of 63°, which is close to the value obtained (around 17-18 %, Fig. 10) for the mixtures that best match its reflectance spectra (olivine-FeS 60:40 to 70:30 vol%, Fig. 13). The reflectance of these mixture are of 0.077 and 0.066 that convert to equivalent albedos of 0.087 and 0.073 using the law derived in Beck et al. (2021), and are slightly higher than the average of B-types in DeMeo and Carry (2013) (0.071 ± 0.033 with a mode at 0.061 ± 0.21).

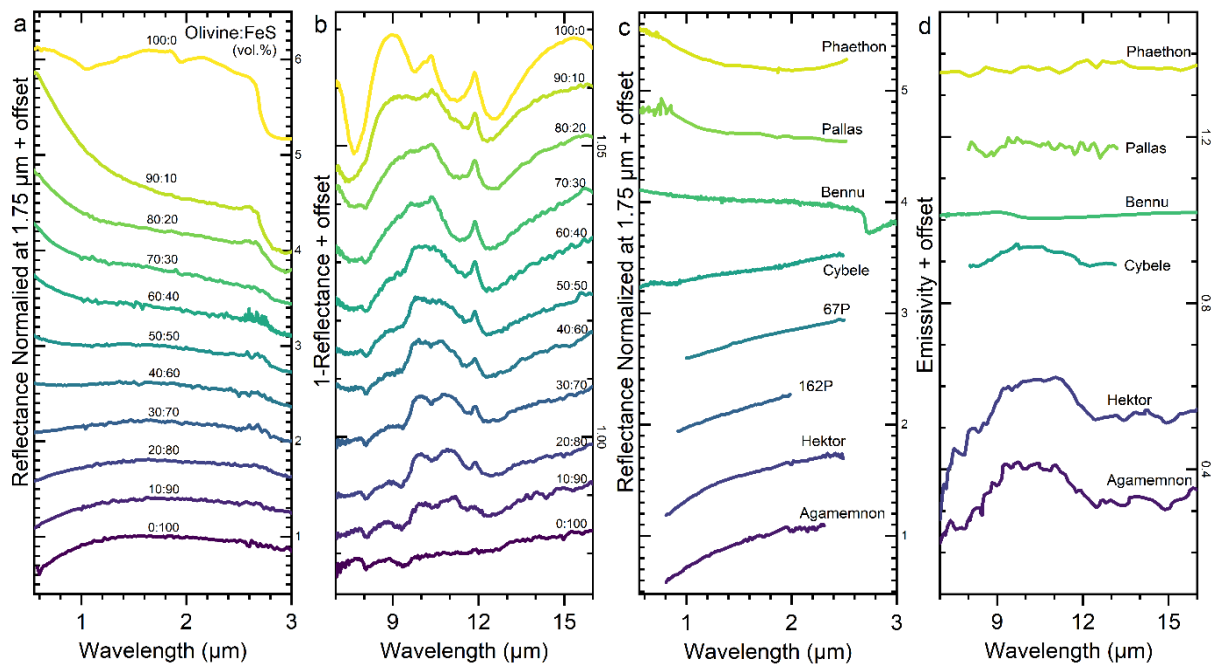

*Figure 15: Comparison of the spectra of mixtures of sub-µm grains of Olivine and FeS (**a, b**), with Vis-NIR reflectance spectra and MIR emission spectra of small bodies (**c, d**). In the mixtures with low proportions of opaques, the Vis-NIR spectral slope is blue due to Rayleigh-like scattering by isolated opaque sub-µm grains. In mixtures with higher proportions of opaques, the Vis-NIR slope becomes redder as the size of the opaque aggregates increases. As the proportion of opaques increases, the silicate grains are de-agglomerated and isolated, de-saturating the silicate absorption bands in the MIR and resulting in the emergence of the 10-µm plateau. Spectra of B-type asteroids (Phaethon, Pallas, Bennu) have a blue Vis-NIR slope and lack a 10-µm plateau, whereas spectra of P/D/X/C-type asteroids (Hektor, Agamemnon) and comets have a Vis-NIR red slope and a 10-µm plateau. The comparison with the mixtures suggests that variations of the degree of dispersion/aggregation of sub-*

*µm grains in the surface material of asteroids and comets is a major parameter explaining the variations of their spectral properties.* Vis-NIR spectra are *from Kareta et al. (2018b) for Phaethon, Rivkin and DeMeo (2019) for Pallas, Hamilton et al. (2019b) for Bennu, the SMASS/MIT database for Cybele, Raponi et al. (2020)* for comet 67P/C.G., *Campins et al. (2006) for comet 162P, Emery and Brown (2003) for Hektor and Agamemnon. MIR spectra are from Lim et al., (2019) for Phaethon, Lim et al. (2005) for Pallas, Hamilton et al. (2019b) for Bennu, Licandro et al. (2011) for Cybele, and Emery et al. (2006) for Hektor and Agamemnon.*

## 5. Conclusion

This work was aimed to test the hypothesis that hyperfine grains control, or have a significant contribution, to the spectral and polarimetric properties of primitive bodies of the Solar system. We prepared and measured the properties of mixtures of bright and opaque materials of sub-micrometric grains at different concentrations. We draw the following observations and conclusions:

- Mixtures of opaques and silicates in the realm of hyperfine grains (< 1 µm) reveal a non-linear behaviour. While the strong darkening effect of very fine opaques mixed with larger translucent grains has been shown in several earlier studies, we show that this effect also occurs when mixing hyperfine grains of opaques with hyperfine grains of silicates. Notably, a small amount of opaque material ($\geq$ 1-5 vol%) is sufficient to mask the absorption bands of silicates in the Vis-NIR, explaining their absence on P/D-type asteroids and comets spectra.

- Relatively low amounts of opaques mixed with the silicates can produce a strong bluing in the visible range. By performing an analysis similar to Brown (2014) we show that this bluing is likely due to a Rayleigh-like scattering in our particulate sample.

- The polarimetric phase curves of these mixtures show a highly non-linear behaviour with increasing fraction of opaques at low phase angles, i.e. in the negative branch. The $|P_{min}|$ value is always higher for the mixtures than for the endmembers. The $|P_{min}|$ value is highest for the mixture of olivine 90 vol% and opaque iron sulfide (FeS) 10 vol% and then progressively decreases with increasing FeS content. The inversion angle decreases from the pure olivine to the olivine 90 % and FeS 10 %, then increases for the 50:50 % mixture, then decreases progressively toward the pure FeS value. At high phase angles, i.e. in the positive branch, the intensity of linear polarization value is roughly correlated with the reflectance.

- Emissivity calculated in the MIR from Kirchoff's law for these hyperfine mixtures can show a 10-µm emissivity peak that resembles the silicate signature observed in transmission (i.e. with no contribution from the real part of the optical constant). We explain this observation by an optical separation of silicate grains by opaque grains, and a diffusion of the MIR light by the opaque grains, enabling photons to escape after some absorption by silicate grains. This effect is similar to observation when mixing silicate with infrared-transparent salt to simulate porosity, as observed in previous works (King et al., 2011; Yang et al., 2013; Izawa et al., 2021). However, while mixtures with KBr are very reflective, mixtures with FeS have a much lower reflectance, compatible with the low reflectance and high emissivity values observed on small bodies.

- The separation of silicate grains, not only by vacuum (elevated micro-rugosity and/or micro-porosity) but also by opaque grains, can explain the peculiar emissivity spectra of P/D-type asteroids and their resemblance to emission spectra from cometary dust tails. Opaque grains (opaque in the visible) in fine-grained mixtures contribute to isolate optically individual silicate grains and reduce optical path length within the silicate phase. These measurements show that the quality of the P/D-type asteroid spectra in terms of the 10-µm emissivity plateau is explained with hyperfine opaques, but not the spectral contrast. The polarimetric properties of some of our MIR spectral analogues are similar (but not strictly identical) to those measured for P/D-type asteroid surfaces.

- Rayleigh-like scattering as observed in our mixture may explain the spectra of B-type asteroids. Our mixtures exhibiting Rayleigh-like scattering have MIR spectra that do not show the silicate 10-µm emission plateau, in agreement with MIR spectra of the B-type Bennu. The positive branch of polarization is similar to that of the B-type Phaethon, but the negative branch is different from those measured for B-type asteroids. Although Bennu appears to be made of rocks and not fine-grained regolith, the presence of dispersed hyperfine grained opaques in these porous rocks could provide an explanation for its blue spectral slope. To improve the analogy with B-types, future studies should investigate how different degrees of consolidation of mixtures of hyperfine grains influence their spectra.

- Our measurements suggest that the apparent correlation of NIR slope with 10-µm plateau observed on P/D/X/C- and B-types (Emery et al., 2011; Marchis et al., 2012; Beck and Poch, 2021) could be due to different degrees of dispersion/aggregation of sub-µm grains of silicate and opaque materials at their surfaces.

Finally, the mixtures we measured that resemble the comets and P/D-type asteroids spectra best contain a large proportion of opaque minerals (50-90 vol%), which is unrealistic for these objects. CP-IDPs whose spectra match very well these objects (Vernazza et al., 2015) contain less than 40 vol% of Fe-Ni sulfides (Bradley, 2014). However and interestingly, CP-IDPs contain a proportion of crystalline silicates (~20-50 vol%, Alexander et al., 2007) similar to the mixtures bearing the best resemblance to the observations (10-50 vol%). In our mixtures, the optical separation of crystalline silicate grains may be quantitatively similar to CP-IDPs, but the compositional and textural parameters controlling the separation are different. Indeed, our mixtures lack important characteristics of CP-IDPs: (1) analogous carbonaceous materials and other components, (2) fluffy microstructures, (3) grain size distributions from 0.1 µm to several micrometres, and cementing semi-continuous matrixes. Some of these characteristics and possibly others (i.e. space weathering) could also explain why the shape and magnitude of the Vis-NIR spectral slope and of the 10-µm feature are different between the samples studied here and the comets and P/D-type asteroids (Figure 13b, Figure 14). Future experiments aiming at understanding the optical properties of primitive small bodies should thus be dedicated to the production and measurement of mixtures having more realistic composition and texture (different grain size distributions, higher micro-porosity etc.).

**Acknowledgements**

This work has been funded by the European Research Council (ERC) under the grant SOLARYS ERC-CoG2017-771691. We acknowledge Nathaniel Findling and Bruno Lanson from ISTerre for the XRD measurements, Frédérique Charlot from the Consortium des Moyens Techniques Communs (CMTC) of University Grenoble Alpes for the SEM images. We acknowledge Laurène Flandinet and Olivier Brissaud from IPAG for their help with the sample preparation protocol and the use the spectroscopic facilities respectively. The contributions of SS, CLP and AP have been carried out within the framework of the NCCR PlanetS supported by the Swiss National Science Foundation under grants 51NF40_182901 and 51NF40_205606. Part of this work has been done during a transnational access visit in the frame of the Europlanet 2020 RI program. Europlanet 2020 RI has received funding from the European Union's Horizon 2020 research and innovation programme under grant agreement No 654208. Support from the Centre national d'étude spatiales (CNES) is also acknowledged. Part of the data shown in this publication were obtained and made available by the MITHNEOS MIT-Hawaii Near-Earth Object Spectroscopic Survey. The IRTF is operated by the University of Hawaii under contract 80HQTR19D0030 with the National Aeronautics and Space Administration. The MIT component of this work is supported by NASA grant 80NSSC18K0849. We are grateful to Pierre Vernazza and an anonymous reviewer for their reviews and suggestions.

**Data availability**

All measured spectra and their associated sample information are freely available through the GhOSST database of the SSHADE infrastructure for solid spectroscopy, supported by the Europlanet 2020-RI program. Direct links to the data are provided in the references (Sultana 2019a, 2019b, 2019c, 2019d). The polarimetric data are also freely available (Sultana et al., 2023).

**Supplementary Figures:**

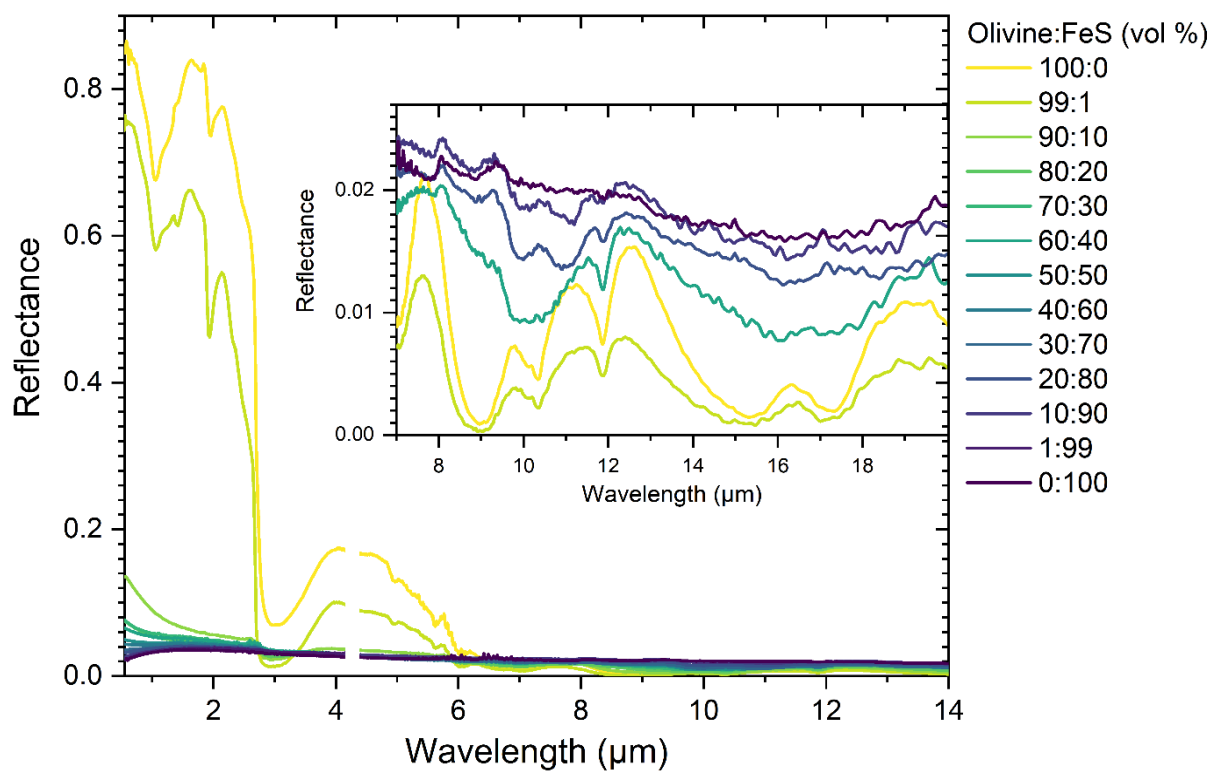

**Supplementary Figure 1:** Reflectance spectra of the mixtures of sub-µm grains of olivine and iron sulfide. From 0.5 to 4.2 µm, the spectra were measured with the SHADOWS goniometer, and from 2.0 to 20 µm the spectra were measured with a FTIR goniometer. To produce this plot, the spectra measured with the FTIR goniometer were scaled on the absolute reflectance level measured by the SHADOWS goniometer at wavelength between 2.5 to 4.0 µm depending on the spectrum.

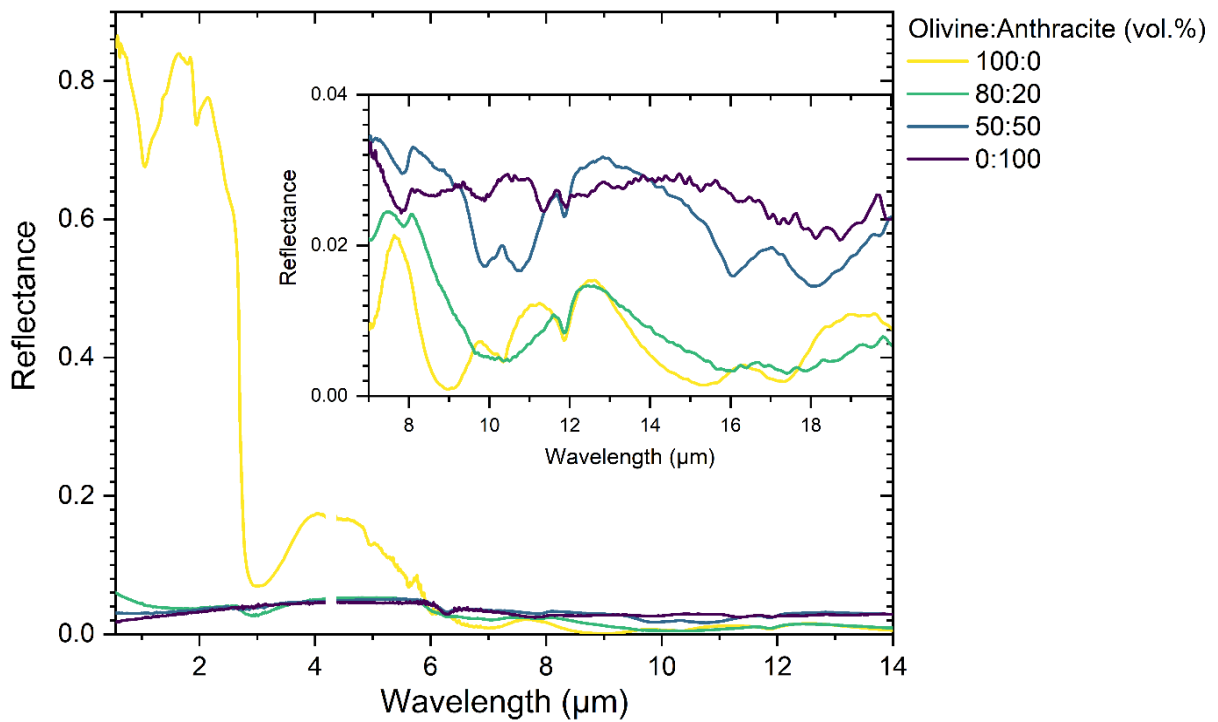

**Supplementary Figure 2:** Reflectance spectra of the mixtures of sub-µm grains of olivine and anthracite. From 0.5 to 4.2 µm, the spectra were measured with the SHADOWS goniometer, and from 2.0 to 20 µm the spectra were measured with a FTIR goniometer. To produce this plot, the spectra measured with the FTIR goniometer were scaled on the absolute reflectance level measured by the SHADOWS goniometer at wavelength between 2.5 to 4.0 µm depending on the spectrum.

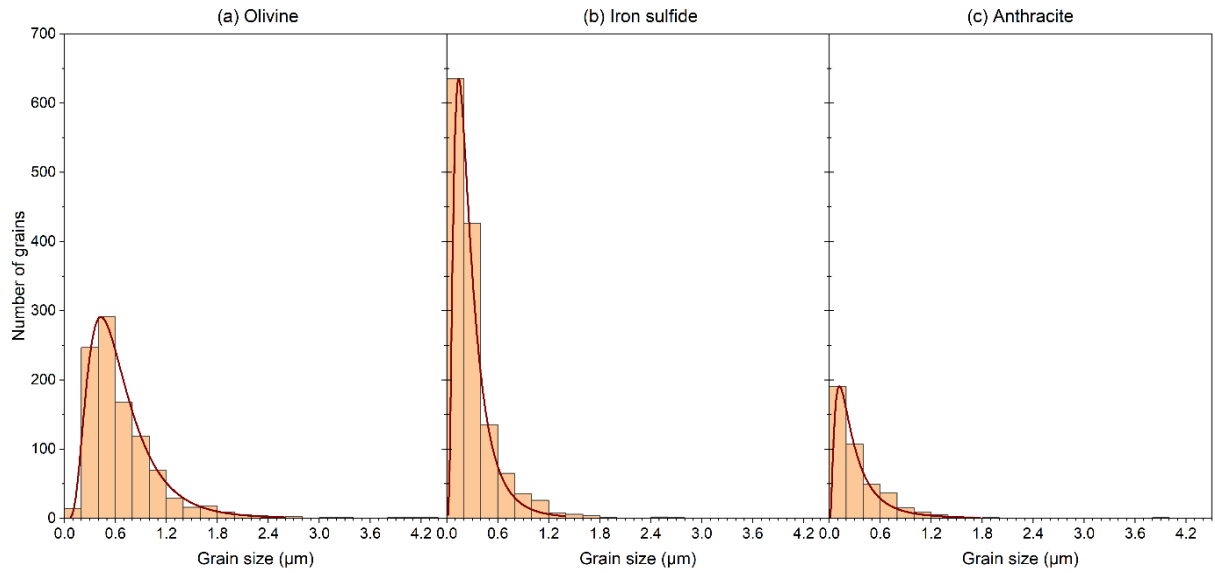

**Supplementary Figure 3:** Size distributions of the powders produced and measured in this study. The red curves are lognormal functions fitting the data. Table 1 shows the mean, standard deviation, median and maximal sizes of the grains.

**References:**

Albiniak, A., Furdin, G., Begin, D., Mareche, J.F., Kaczmarczyk, J., Broniek, E., 1996. Exfoliation and textural modification of anthracites. Carbon 34, 1329–1334. https://doi.org/10.1016/S0008-6223(96)00066-8

Alexander, C., Boss, A., Keller, L., Nuth, J., Weinberger, A., 2007. Astronomical and meteoritic evidence for the nature of interstellar dust and its processing in protoplanetary disks. Protostars and planets V 801–813.

Arnold, G., Wagner, C., 1988. Grain-size influence on the mid-infrared spectra of the minerals. Earth Moon Planet 41, 163–171. https://doi.org/10.1007/BF00056401

Bagnulo, S., Belskaya, I., Stinson, A., Christou, A., Borisov, G.B., 2016. Broadband linear polarization of Jupiter Trojans. Astronomy & Astrophysics 585, A122. https://doi.org/10.1051/0004-6361/201526889

Bagnulo, S., Cellino, A., Sterzik, M.F., 2015. Linear spectropolarimetry: a new diagnostic tool for the classification and characterization of asteroids. Mon Not R Astron Soc Lett 446, L11–L15. https://doi.org/10.1093/mnrasl/slu154

Beck, P., Eschrig, J., Potin, S., Prestgard, T., Bonal, L., Quirico, E., Schmitt, B., 2021. "Water" abundance at the surface of C-complex main-belt asteroids. Icarus 357, 114125. https://doi.org/10.1016/j.icarus.2020.114125

Beck, P., Poch, O., 2021. Origins of colors variability among C-cluster main-belt asteroids. Icarus 365, 114494. https://doi.org/10.1016/j.icarus.2021.114494

Belskaya, I., Cellino, A., Levasseur-Regourd, A.-C., Bagnulo, S., 2019. Optical Polarimetry of Small Solar System Bodies: From Asteroids to Debris Disks, in: Mignani, R., Shearer, A., Słowikowska, A., Zane, S. (Eds.), Astronomical Polarisation from the Infrared to Gamma Rays, Astrophysics and Space Science Library. Springer International Publishing, Cham, pp. 223–246. https://doi.org/10.1007/978-3-030-19715-5_9

Belskaya, I.N., Fornasier, S., Tozzi, G.P., Gil-Hutton, R., Cellino, A., Antonyuk, K., Krugly, Yu.N., Dovgopol, A.N., Faggi, S., 2017. Refining the asteroid taxonomy by polarimetric observations. Icarus 284, 30–42. https://doi.org/10.1016/j.icarus.2016.11.003

Belskaya, I.N., Shkuratov, Yu.G., Efimov, Yu.S., Shakhovskoy, N.M., Gil-Hutton, R., Cellino, A., Zubko, E.S., Ovcharenko, A.A., Bondarenko, S.Yu., Shevchenko, V.G., Fornasier, S., Barbieri, C., 2005. The F-type asteroids with small inversion angles of polarization. Icarus 178, 213–221. https://doi.org/10.1016/j.icarus.2005.04.015

Blokh, A.G., Burak, D., 1972. Determination of the IR refractive index and absorption coefficient of solid fuels. Journal of engineering physics 23, 7–9.

Bradley, J.P., 2014. 1.8 - Early Solar Nebula Grains – Interplanetary Dust Particles, in: Holland, H.D., Turekian, K.K. (Eds.), Treatise on Geochemistry (Second Edition). Elsevier, Oxford, pp. 287–308. https://doi.org/10.1016/B978-0-08-095975-7.00114-5

Brown, A.J., 2014. Spectral bluing induced by small particles under the Mie and Rayleigh regimes. Icarus 239, 85–95. https://doi.org/10.1016/j.icarus.2014.05.042

Brownlee, D.E., 2014. 2.13 - Comets, in: Holland, H.D., Turekian, K.K. (Eds.), Treatise on Geochemistry (Second Edition). Elsevier, Oxford, pp. 335–363. https://doi.org/10.1016/B978-0-08-095975-7.00128-5

Brunetto, R., Borg, J., Dartois, E., Rietmeijer, F.J.M., Grossemy, F., Sandt, C., Le Sergeant d'Hendecourt, L., Rotundi, A., Dumas, P., Djouadi, Z., Jamme, F., 2011. Mid-IR, Far-IR, Raman micro-spectroscopy, and FESEM–EDX study of IDP L2021C5: Clues to its origin. Icarus 212, 896–910. https://doi.org/10.1016/j.icarus.2011.01.038

Campins, H., Ziffer, J., Licandro, J., Pinilla-Alonso, N., Fernández, Y., de León, J., Mothé-Diniz, T., Binzel, R.P., 2006. Nuclear Spectra of Comet 162P/Siding Spring (2004 TU12). The Astronomical Journal 132, 1346–1353. https://doi.org/10.1086/506253

Capaccioni, F., Coradini, A., Filacchione, G., Erard, S., Arnold, G., Drossart, P., De Sanctis, M.C., Bockelee-Morvan, D., Capria, M.T., Tosi, F., Leyrat, C., Schmitt, B., Quirico, E., Cerroni, P., Mennella, V., Raponi, A., Ciarniello, M., McCord, T., Moroz, L., Palomba, E., Ammannito, E., Barucci, M.A., Bellucci, G., Benkhoff, J., Bibring, J.P., Blanco, A., Blecka, M., Carlson, R., Carsenty, U., Colangeli, L., Combes, M., Combi, M., Crovisier, J., Encrenaz, T., Federico, C., Fink, U., Fonti, S., Ip, W.H., Irwin, P., Jaumann, R., Kuehrt, E., Langevin, Y., Magni, G., Mottola, S., Orofino, V., Palumbo, P., Piccioni, G., Schade, U., Taylor, F., Tiphene, D., Tozzi, G.P., Beck, P., Biver, N., Bonal, L., Combe, J.-P., Despan, D., Flamini, E., Fornasier, S., Frigeri, A., Grassi, D., Gudipati, M., Longobardo, A., Markus, K., Merlin, F., Orosei, R., Rinaldi, G., Stephan, K., Cartacci, M., Cicchetti, A., Giuppi, S., Hello, Y., Henry, F., Jacquinod, S., Noschese, R., Peter, G., Politi, R., Reess, J.M., Semery, A., 2015. The organic-rich surface of comet 67P/Churyumov-Gerasimenko as seen by VIRTIS/Rosetta. Science 347, aaa0628–aaa0628. https://doi.org/10.1126/science.aaa0628

Cellino, A., Bagnulo, S., Gil-Hutton, R., Tanga, P., Cañada-Assandri, M., Tedesco, E.F., 2015. On the calibration of the relation between geometric albedo and polarimetric properties for the asteroids★. Monthly Notices of the Royal Astronomical Society 451, 3473–3488. https://doi.org/10.1093/mnras/stv1188

Clark, B.E., Binzel, R.P., Howell, E.S., Cloutis, E.A., Ockert-Bell, M., Christensen, P., Barucci, M.A., DeMeo, F., Lauretta, D.S., Connolly, H., Soderberg, A., Hergenrother, C., Lim, L., Emery, J., Mueller, M., 2011. Asteroid (101955) 1999 RQ36: Spectroscopy from 0.4 to 2.4µm and meteorite analogs. Icarus 216, 462–475. https://doi.org/10.1016/j.icarus.2011.08.021

Clark, R.N., Curchin, J.M., Jaumann, R., Cruikshank, D.P., Brown, R.H., Hoefen, T.M., Stephan, K., Moore, J.M., Buratti, B.J., Baines, K.H., Nicholson, P.D., Nelson, R.M., 2008. Compositional mapping of Saturn's satellite Dione with Cassini VIMS and implications of dark material in the Saturn system. Icarus, Saturn's Icy Satellites from Cassini 193, 372–386. https://doi.org/10.1016/j.icarus.2007.08.035

Cloutis, E.A., Pietrasz, V.B., Kiddell, C., Izawa, M.R.M., Vernazza, P., Burbine, T.H., DeMeo, F., Tait, K.T., Bell, J.F., Mann, P., Applin, D.M., Reddy, V., 2018. Spectral reflectance "deconstruction" of the Murchison CM2 carbonaceous chondrite and implications for spectroscopic investigations of dark asteroids. Icarus 305, 203–224. https://doi.org/10.1016/j.icarus.2018.01.015

Dai, Z.R., Bradley, J.P., 2001. Iron-nickel sulfides in anhydrous interplanetary dust particles. Geochimica et Cosmochimica Acta 65, 3601–3612. https://doi.org/10.1016/S0016-7037(01)00692-5

Delsanti, A.C., Boehnhardt, H., Barrera, L., Meech, K.J., Sekiguchi, T., Hainaut, O.R., 2001. BVRI Photometry of 27 Kuiper Belt Objects with ESO/Very Large Telescope. A&A 380, 347–358. https://doi.org/10.1051/0004-6361:20011432

DeMeo, F.E., Binzel, R.P., Slivan, S.M., Bus, S.J., 2009. An extension of the Bus asteroid taxonomy into the near-infrared. Icarus 202, 160–180. https://doi.org/10.1016/j.icarus.2009.02.005

DeMeo, F.E., Carry, B., 2013. The taxonomic distribution of asteroids from multi-filter all-sky photometric surveys. Icarus 226, 723–741. https://doi.org/10.1016/j.icarus.2013.06.027

Dollfus, A., Wolff, M., Geake, J.E., Lupishko, D.F., Dougherty, L.M., 1989. Photopolarimetry of asteroids 594–616.

Emery, J.P., Brown, R.H., 2003. Constraints on the surface composition of Trojan asteroids from near-infrared (0.8–4.0 μm) spectroscopy. Icarus 164, 104–121. https://doi.org/10.1016/S0019-1035(03)00143-X

Emery, J.P., Burr, D.M., Cruikshank, D.P., 2011. Near-infrared Spectroscopy of Trojan Asteroids: Evidence for Two Compositional Groups. The Astronomical Journal 141, 25. https://doi.org/10.1088/0004-6256/141/1/25

Emery, J.P., Cruikshank, D.P., Van Cleve, J., 2006. Thermal emission spectroscopy (5.2–38 μm) of three Trojan asteroids with the Spitzer Space Telescope: Detection of fine-grained silicates. Icarus, Results from the Mars Express ASPERA-3 Investigation 182, 496–512. https://doi.org/10.1016/j.icarus.2006.01.011

Fornasier, S., Hasselmann, P.H., Barucci, M.A., Feller, C., Besse, S., Leyrat, C., Lara, L., Gutierrez, P.J., Oklay, N., Tubiana, C., Scholten, F., Sierks, H., Barbieri, C., Lamy, P.L., Rodrigo, R., Koschny, D., Rickman, H., Keller, H.U., Agarwal, J., A'Hearn, M.F., Bertaux, J.-L., Bertini, I., Cremonese, G., Da Deppo, V., Davidsson, B., Debei, S., De Cecco, M., Fulle, M., Groussin, O., Güttler, C., Hviid, S.F., Ip, W., Jorda, L., Knollenberg, J., Kovacs, G., Kramm, R., Kührt, E., Küppers, M., La Forgia, F., Lazzarin, M., Moreno, J.J.L., Marzari, F., Matz, K.-D., Michalik, H., Moreno, F., Mottola, S., Naletto, G., Pajola, M., Pommerol, A., Preusker, F., Shi, X., Snodgrass, C., Thomas, N., Vincent, J.-B., 2015. Spectrophotometric properties of the nucleus of comet 67P/Churyumov-Gerasimenko from the OSIRIS instrument onboard the ROSETTA spacecraft 30. https://doi.org/10.1051/0004-6361/201525901

Gradie, J., Tedesco, E., 1982. Compositional Structure of the Asteroid Belt. Science 216, 1405–1407. https://doi.org/10.1126/science.216.4553.1405

Güttler, C., Mannel, T., Rotundi, A., Merouane, S., Fulle, M., Bockelée-Morvan, D., Lasue, J., et al., 2019. Synthesis of the morphological description of cometary dust at comet 67P/Churyumov-Gerasimenko. A&A. https://doi.org/10.1051/0004-6361/201834751

Hamilton, V.E., Simon, A.A., Christensen, P.R., Reuter, D.C., Clark, B.E., Barucci, M.A., Bowles, N.E., Boynton, W. V., Brucato, J.R., Cloutis, E.A., Connolly, H.C., Donaldson Hanna, K.L., Emery, J.P., Enos, H.L., Fornasier, S., Haberle, C.W., Hanna, R.D., Howell, E.S., Kaplan, H.H., Keller, L.P., Lantz, C., Li, J.Y., Lim, L.F., McCoy, T.J., Merlin, F., Nolan, M.C., Praet, A., Rozitis, B., Sandford, S.A., Schrader, D.L., Thomas, C.A., Zou, X.D., Lauretta, D.S., Highsmith, D.E., Small, J., Vokrouhlický, D., Bowles, N.E., Brown, E., Donaldson Hanna, K.L., Warren, T., Brunet, C., Chicoine, R.A., Desjardins, S., Gaudreau, D., Haltigin, T., Millington-Veloza, S., Rubi, A., Aponte, J., Gorius, N., Lunsford, A., Allen, B., Grindlay, J., Guevel, D., Hoak, D., Hong, J., Schrader, D.L., Bayron, J., Golubov, O., Sánchez, P., Stromberg, J., Hirabayashi, M., Hartzell, C.M., Oliver, S., Rascon, M., Harch, A., Joseph, J., Squyres, S., Richardson, D., Emery, J.P., McGraw, L., Ghent, R., Binzel, R.P., Asad, M.M.A., Johnson, C.L., Philpott, L., Susorney, H.C.M., Cloutis, E.A., Hanna, R.D., Connolly, H.C., Ciceri, F., Hildebrand, A.R., Ibrahim, E.M., Breitenfeld, L., Glotch, T., Rogers, A.D., Clark, B.E., Ferrone, S., Thomas, C.A., Campins, H., Fernandez, Y., Chang, W., Cheuvront, A., Trang, D., Tachibana, S., Yurimoto, H., Brucato, J.R., Poggiali, G., Pajola, M., Dotto, E., Mazzotta Epifani, E., Crombie, M.K., Lantz, C., Izawa, M.R.M., de Leon, J., Licandro, J., Garcia, J.L.R., Clemett, S., Thomas-Keprta, K., Van wal, S., Yoshikawa, M., Bellerose, J., Bhaskaran, S., Boyles, C., Chesley, S.R., Elder, C.M., Farnocchia, D., Harbison, A., Kennedy, B., Knight, A., Martinez-Vlasoff, N., Mastrodemos, N., McElrath, T., Owen, W., Park, R., Rush, B., Swanson, L., Takahashi, Y., Velez, D., Yetter, K., Thayer, C., Adam, C., Antreasian, P., Bauman, J., Bryan, C., Carcich, B., Corvin, M., Geeraert, J., Hoffman, J., Leonard, J.M., Lessac-Chenen, E., Levine, A., McAdams,

J., McCarthy, L., Nelson, D., Page, B., Pelgrift, J., Sahr, E., Stakkestad, K., Stanbridge, D., Wibben, D., Williams, B., Williams, K., Wolff, P., Hayne, P., Kubitschek, D., Deshapriya, J.D.P., Fornasier, S., Fulchignoni, M., Hasselmann, P., Merlin, F., Praet, A., Bierhaus, E.B., Billett, O., Boggs, A., Buck, B., Carlson-Kelly, S., Cerna, J., Chaffin, K., Church, E., Coltrin, M., Daly, J., Deguzman, A., Dubisher, R., Eckart, D., Ellis, D., Falkenstern, P., Fisher, A., Fisher, M.E., Fleming, P., Fortney, K., Francis, S., Freund, S., Gonzales, S., Haas, P., Hasten, A., Hauf, D., Hilbert, A., Howell, D., Jaen, F., Jayakody, N., Jenkins, M., Johnson, K., Lefevre, M., Ma, H., Mario, C., Martin, K., May, C., McGee, M., Miller, B., Miller, C., Miller, G., Mirfakhrai, A., Muhle, E., Norman, C., Olds, R., Parish, C., Ryle, M., Schmitzer, M., Sherman, P., Skeen, M., Susak, M., Sutter, B., Tran, Q., Welch, C., Witherspoon, R., Wood, J., Zareski, J., Arvizu-Jakubicki, M., Asphaug, E., Audi, E., Ballouz, R.L., Bandrowski, R., Becker, K.J., Becker, T.L., Bendall, S., Bennett, C.A., Bloomenthal, H., Blum, D., Boynton, W.V., Brodbeck, J., Burke, K.N., Chojnacki, M., Colpo, A., Contreras, J., Cutts, J., Drouet d'Aubigny, C.Y., Dean, D., DellaGiustina, D.N., Diallo, B., Drinnon, D., Drozd, K., Enos, H.L., Enos, R., Fellows, C., Ferro, T., Fisher, M.R., Fitzgibbon, G., Fitzgibbon, M., Forelli, J., Forrester, T., Galinsky, I., Garcia, R., Gardner, A., Golish, D.R., Habib, N., Hamara, D., Hammond, D., Hanley, K., Harshman, K., Hergenrother, C.W., Herzog, K., Hill, D., Hoekenga, C., Hooven, S., Howell, E.S., Huettner, E., Janakus, A., Jones, J., Kareta, T.R., Kidd, J., Kingsbury, K., Balram-Knutson, S.S., Koelbel, L., Kreiner, J., Lambert, D., Lauretta, D.S., Lewin, C., Lovelace, B., Loveridge, M., Lujan, M., Maleszewski, C.K., Malhotra, R., Marchese, K., McDonough, E., Mogk, N., Morrison, V., Morton, E., Munoz, R., Nelson, J., Nolan, M.C., Padilla, J., Pennington, R., Polit, A., Ramos, N., Reddy, V., Riehl, M., Rizk, B., Roper, H.L., Salazar, S., Schwartz, S.R., Selznick, S., Shultz, N., Smith, P.H., Stewart, S., Sutton, S., Swindle, T., Tang, Y.H., Westermann, M., Wolner, C.W.V., Worden, D., Zega, T., Zeszut, Z., Bjurstrom, A., Bloomquist, L., Dickinson, C., Keates, E., Liang, J., Nifo, V., Taylor, A., Teti, F., Caplinger, M., Bowles, H., Carter, S., Dickenshied, S., Doerres, D., Fisher, T., Hagee, W., Hill, J., Miner, M., Noss, D., Piacentine, N., Smith, M., Toland, A., Wren, P., Bernacki, M., Munoz, D.P., Watanabe, S.I., Sandford, S.A., Aqueche, A., Ashman, B., Barker, M., Bartels, A., Berry, K., Bos, B., Burns, R., Calloway, A., Carpenter, R., Castro, N., Cosentino, R., Donaldson, J., Dworkin, J.P., Elsila Cook, J., Emr, C., Everett, D., Fennell, D., Fleshman, K., Folta, D., Gallagher, D., Garvin, J., Getzandanner, K., Glavin, D., Hull, S., Hyde, K., Ido, H., Ingegneri, A., Jones, N., Kaotira, P., Lim, L.F., Liounis, A., Lorentson, C., Lorenz, D., Lyzhoft, J., Mazarico, E.M., Mink, R., Moore, W., Moreau, M., Mullen, S., Nagy, J., Neumann, G., Nuth, J., Poland, D., Reuter, D.C., Rhoads, L., Rieger, S., Rowlands, D., Sallitt, D., Scroggins, A., Shaw, G., Simon, A.A., Swenson, J., Vasudeva, P., Wasser, M., Zellar, R., Grossman, J., Johnston, G., Morris, M., Wendel, J., Burton, A., Keller, L.P., McNamara, L., Messenger, S., Nakamura-Messenger, K., Nguyen, A., Righter, K., Queen, E., Bellamy, K., Dill, K., Gardner, S., Giuntini, M., Key, B., Kissell, J., Patterson, D., Vaughan, D., Wright, B., Gaskell, R.W., Le Corre, L., Li, J.Y., Molaro, J.L., Palmer, E.E., Siegler, M.A., Tricarico, P., Weirich, J.R., Zou, X.D., Ireland, T., Tait, K., Bland, P., Anwar, S., Bojorquez-Murphy, N., Christensen, P.R., Haberle, C.W., Mehall, G., Rios, K., Franchi, I., Rozitis, B., Beddingfield, C.B., Marshall, J., Brack, D.N., French, A.S., McMahon, J.W., Scheeres, D.J., Jawin, E.R., McCoy, T.J., Russell, S., Killgore, M., Bottke, W.F., Hamilton, V.E., Kaplan, H.H., Walsh, K.J., Bandfield, J.L., Clark, B.C., Chodas, M., Lambert, M., Masterson, R.A., Daly, M.G., Freemantle, J., Seabrook, J.A., Barnouin, O.S., Craft, K., Daly, R.T., Ernst, C.,

Espiritu, R.C., Holdridge, M., Jones, M., Nair, A.H., Nguyen, L., Peachey, J., Perry, M.E., Plescia, J., Roberts, J.H., Steele, R., Turner, R., Backer, J., Edmundson, K., Mapel, J., Milazzo, M., Sides, S., Manzoni, C., May, B., Delbo', M., Libourel, G., Michel, P., Ryan, A., Thuillet, F., Marty, B., 2019a. Evidence for widespread hydrated minerals on asteroid (101955) Bennu. Nature Astronomy 3, 332–340. https://doi.org/10.1038/s41550-019-0722-2

Hamilton, V.E., Simon, A.A., Christensen, P.R., Reuter, D.C., Clark, B.E., Barucci, M.A., Bowles, N.E., Boynton, W.V., Brucato, J.R., Cloutis, E.A., Connolly, H.C., Hanna, K.L.D., Emery, J.P., Enos, H.L., Fornasier, S., Haberle, C.W., Hanna, R.D., Howell, E.S., Kaplan, H.H., Keller, L.P., Lantz, C., Li, J.-Y., Lim, L.F., McCoy, T.J., Merlin, F., Nolan, M.C., Praet, A., Rozitis, B., Sandford, S.A., Schrader, D.L., Thomas, C.A., Zou, X.-D., Lauretta, D.S., 2019b. Evidence for widespread hydrated minerals on asteroid (101955) Bennu. Nature Astronomy 1. https://doi.org/10.1038/s41550-019-0722-2

Hanuš, J., Vokrouhlický, D., Delbo', M., Farnocchia, D., Polishook, D., Pravec, P., Hornoch, K., Kučáková, H., Kušnirák, P., Stephens, R., Warner, B., 2018. (3200) Phaethon: Bulk density from Yarkovsky drift detection. A&A 620, L8. https://doi.org/10.1051/0004-6361/201834228

Hapke, B., 2012. Theory of reflectance and emittance spectroscopy. Cambridge university press.

Hapke, B., 2001. Space weathering from Mercury to the asteroid belt. Journal of Geophysical Research: Planets 106, 10039–10073. https://doi.org/10.1029/2000JE001338

Howard, K.T., Benedix, G.K., Bland, P.A., Cressey, G., 2011. Modal mineralogy of CM chondrites by X-ray diffraction (PSD-XRD): Part 2. Degree, nature and settings of aqueous alteration. Geochimica et Cosmochimica Acta 75, 2735–2751. https://doi.org/10.1016/j.gca.2011.02.021

Hunt, G.R., Logan, L.M., 1972. Variation of Single Particle Mid-Infrared Emission Spectrum with Particle Size. Appl. Opt. 11, 142. https://doi.org/10.1364/AO.11.000142

Hunt, G.R., Vincent, R.K., 1968. The behavior of spectral features in the infrared emission from particulate surfaces of various grain sizes. Journal of Geophysical Research (1896-1977) 73, 6039–6046. https://doi.org/10.1029/JB073i018p06039

Izawa, M.R.M., King, P.L., Vernazza, P., Berger, J.A., McCutcheon, W.A., 2021. Salt – A critical material to consider when exploring the solar system. Icarus 359, 114328. https://doi.org/10.1016/j.icarus.2021.114328

Johnson, T.V., Fanale, F.P., 1973. Optical properties of carbonaceous chondrites and their relationship to asteroids. Journal of Geophysical Research (1896-1977) 78, 8507–8518. https://doi.org/10.1029/JB078i035p08507

Kareta, T., Reddy, V., Hergenrother, C., Lauretta, D.S., Arai, T., Takir, D., Sanchez, J., Hanuš, J., 2018a. Rotationally Resolved Spectroscopic Characterization of Near-Earth Object (3200) Phaethon. The Astronomical Journal 156, 287. https://doi.org/10.3847/1538-3881/aaeb8a

Kareta, T., Reddy, V., Hergenrother, C., Lauretta, D.S., Arai, T., Takir, D., Sanchez, J., Hanuš, J., 2018b. Rotationally Resolved Spectroscopic Characterization of Near-Earth Object (3200) Phaethon. AJ 156, 287. https://doi.org/10.3847/1538-3881/aaeb8a

Kelley, M.S.P., Woodward, C.E., Gehrz, R.D., Reach, W.T., Harker, D.E., 2017. Mid-infrared spectra of comet nuclei. Icarus 284, 344–358. https://doi.org/10.1016/j.icarus.2016.11.029

King, P.L., Izawa, M.R.M., Vernazza, P., McCutcheon, W.A., Berger, J.A., Dunn, T., 2011. Salt — A Critical Material to Consider when Exploring the Solar System. Presented at the 42nd Annual Lunar and Planetary Science Conference, The Woodlands, Texas, p. 1985.

Lantz, C., Brunetto, R., Barucci, M.A., Fornasier, S., Baklouti, D., Bourçois, J., Godard, M., 2017. Ion irradiation of carbonaceous chondrites: A new view of space weathering on primitive asteroids. Icarus 285, 43–57. https://doi.org/10.1016/j.icarus.2016.12.019

Le Bras, A., Erard, S., 2003. Reflectance spectra of regolith analogs in the mid-infrared: effects of grain size. Planetary and Space Science 51, 281–294. https://doi.org/10.1016/S0032-0633(03)00017-5

Licandro, J., Campins, H., Kelley, M., Hargrove, K., Pinilla-Alonso, N., Cruikshank, D., Rivkin, A.S., Emery, J., 2011. (65) Cybele: detection of small silicate grains, water-ice, and organics. Astronomy & Astrophysics 525, A34. https://doi.org/10.1051/0004-6361/201015339

Licandro, J., Hargrove, K., Kelley, M., Campins, H., Ziffer, J., Alí-Lagoa, V., Fernández, Y., Rivkin, A., 2012. 5–14 $\mu$ m *Spitzer* spectra of Themis family asteroids. Astronomy & Astrophysics 537, A73. https://doi.org/10.1051/0004-6361/201118142

Lim, L.F., Hamilton, V.E., Christensen, P.R., Simon, A.A., Reuter, D.C., Emery, J.P., Rozitis, B., Antonella Barucci, M., Campins, H., Clark, B.E., Delbo, M., Licandro, J., Hanna, R.D., Howell, E.S., Lauretta, D.S., 2019. Main-Belt Infrared Spectral Analogues for (101955) Bennu: AKARI and Spitzer IRS Asteroid Spectra. Presented at the EPSC-DPS Joint Meeting 2019, Geneva, p. EPSC-DPS2019-312.

Lim, L.F., McConnochie, T.H., Bell, J.F., Hayward, T.L., 2005. Thermal infrared (8–13 μm) spectra of 29 asteroids: the Cornell Mid-Infrared Asteroid Spectroscopy (MIDAS) Survey. Icarus 173, 385–408. https://doi.org/10.1016/j.icarus.2004.08.005

Loeffler, M.J., Prince, B.S., 2022. A possible explanation for the blue spectral slope observed on B-type asteroids. Icarus 376, 114881. https://doi.org/10.1016/j.icarus.2022.114881

Mannel, T., Bentley, M.S., Boakes, P.D., Jeszenszky, H., Ehrenfreund, P., Engrand, C., Koeberl, C., Levasseur-Regourd, A.C., Romstedt, J., Schmied, R., Torkar, K., Weber, I., 2019. Dust of comet 67P/Churyumov-Gerasimenko collected by Rosetta/MIDAS: classification and extension to the nanometre scale. A&A. https://doi.org/10.1051/0004-6361/201834851

Marchis, F., Enriquez, J.E., Emery, J.P., Mueller, M., Baek, M., Pollock, J., Assafin, M., Vieira Martins, R., Berthier, J., Vachier, F., Cruikshank, D.P., Lim, L.F., Reichart, D.E., Ivarsen, K.M., Haislip, J.B., LaCluyze, A.P., 2012. Multiple asteroid systems: Dimensions and thermal properties from Spitzer Space Telescope and ground-based observations. Icarus 221, 1130–1161. https://doi.org/10.1016/j.icarus.2012.09.013

Martin, A.C., Emery, J.P., Loeffler, M.J., 2022. Spectral effects of regolith porosity in the mid-IR – Forsteritic olivine. Icarus 378, 114921. https://doi.org/10.1016/j.icarus.2022.114921

Masiero, J.R., Wright, E.L., Mainzer, A.K., 2019. Thermophysical Modeling of NEOWISE Observations of DESTINY+ Targets Phaethon and 2005 UD. AJ 158, 97. https://doi.org/10.3847/1538-3881/ab31a6

Matsuoka, M., Nakamura, T., Hiroi, T., Okumura, S., Sasaki, S., 2020. Space Weathering Simulation with Low-energy Laser Irradiation of Murchison CM Chondrite for Reproducing Micrometeoroid Bombardments on C-type Asteroids. ApJL 890, L23. https://doi.org/10.3847/2041-8213/ab72a4

Maturilli, A., Helbert, J., Arnold, G., 2019. The newly improved set-up at the Planetary Spectroscopy Laboratory (PSL) 2. https://doi.org/10.1117/12.2529266

McAdam, M., Mommert, M., Trilling, D., 2018. Archived Spitzer Observations of (3200) Phaethon: an aqueously altered asteroid. 50, 312.01.

Moersch, J.E., Christensen, P.R., 1995. Thermal emission from particulate surfaces: A comparison of scattering models with measured spectra. Journal of Geophysical Research: Planets 100, 7465–7477. https://doi.org/10.1029/94JE03330

Morbidelli, A., Levison, H.F., Tsiganis, K., Gomes, R., 2005. Chaotic capture of Jupiter's Trojan asteroids in the early Solar System. Nature 435, 462–465. https://doi.org/10.1038/nature03540

Mustard, J.F., Glotch, T., 2020. Theory of Reflectance and Emittance Spectroscopy of Geologic Materials in the Visible and Infrared Regions, in: Bishop, J.L., Bell III, J.F., Bell, J., Moersch, J.E. (Eds.), Remote Compositional Analysis: Techniques for Understanding Spectroscopy, Mineralogy, and Geochemistry of Planetary Surfaces. Cambridge University Press.

Mustard, J.F., Hays, J.E., 1997. Effects of Hyperfine Particles on Reflectance Spectra from 0.3 to 25 μm. Icarus 125, 145–163. https://doi.org/10.1006/icar.1996.5583

Noble, S.K., Pieters, C.M., Keller, L.P., 2007. An experimental approach to understanding the optical effects of space weathering. Icarus 192, 629–642. https://doi.org/10.1016/j.icarus.2007.07.021

Pieters, C.M., Noble, S.K., 2016. Space weathering on airless bodies. Journal of Geophysical Research: Planets 121, 1865–1884. https://doi.org/10.1002/2016JE005128

Poch, O., Cerubini, R., Pommerol, A., Jost, B., Thomas, N., 2018. Polarimetry of Water Ice Particles Providing Insights on Grain Size and Degree of Sintering on Icy Planetary Surfaces. Journal of Geophysical Research: Planets 123, 2564–2584. https://doi.org/10.1029/2018JE005753

Poch, O., Istiqomah, I., Quirico, E., Beck, P., Schmitt, B., Theulé, P., Faure, A., Hily-Blant, P., Bonal, L., Raponi, A., Ciarniello, M., Rousseau, B., Potin, S., Brissaud, O., Flandinet, L., Filacchione, G., Pommerol, A., Thomas, N., Kappel, D., Mennella, V., Moroz, L., Vinogradoff, V., Arnold, G., Erard, S., Bockelée-Morvan, D., Leyrat, C., Capaccioni, F., Sanctis, M.C.D., Longobardo, A., Mancarella, F., Palomba, E., Tosi, F., 2020. Ammonium salts are a reservoir of nitrogen on a cometary nucleus and possibly on some asteroids. Science 367. https://doi.org/10.1126/science.aaw7462

Poch, O., Pommerol, A., Jost, B., Carrasco, N., Szopa, C., Thomas, N., 2016. Sublimation of water ice mixed with silicates and tholins: Evolution of surface texture and reflectance spectra, with implications for comets. Icarus 267, 154–173. https://doi.org/10.1016/j.icarus.2015.12.017

Potin, S., Brissaud, O., Beck, P., Schmitt, B., Magnard, Y., Correia, J.-J., Rabou, P., Jocou, L., 2018. SHADOWS: a spectro-gonio radiometer for bidirectional reflectance studies of dark meteorites and terrestrial analogs: design, calibrations, and performances on challenging surfaces. Applied Optics 57, 8279. https://doi.org/10.1364/AO.57.008279

Price, M.C., Kearsley, A.T., Burchell, M.J., Hörz, F., Borg, J., Bridges, J.C., Cole, M.J., Floss, C., Graham, G., Green, S.F., Hoppe, P., Leroux, H., Marhas, K.K., Park, N., Stroud, R., Stadermann, F.J., Telisch, N., Wozniakiewicz, P.J., 2010. Comet 81P/Wild 2: The size distribution of finer (sub-10 μm) dust collected by the Stardust spacecraft. Meteoritics & Planetary Science 45, 1409–1428. https://doi.org/10.1111/j.1945-5100.2010.01104.x

Querry, M.R., 1985. Optical Constants, Report No. AD-A158 623. Crdc CR-85034, 1–413.

Quirico, E., Moroz, L.V., Schmitt, B., Arnold, G., Faure, M., Beck, P., Bonal, L., Ciarniello, M., Capaccioni, F., Filacchione, G., Erard, S., Leyrat, C., Bockelée-Morvan, D., Zinzi, A., Palomba, E., Drossart, P., Tosi, F., Capria, M.T., De Sanctis, M.C., Raponi, A., Fonti, S., Mancarella, F., Orofino, V., Barucci, A., Blecka, M.I., Carlson, R., Despan, D., Faure, A., Fornasier, S., Gudipati, M.S., Longobardo, A., Markus, K., Mennella, V., Merlin, F., Piccioni, G., Rousseau, B., Taylor, F., 2016. Refractory and semi-volatile organics at the surface of comet 67P/Churyumov-Gerasimenko: Insights from the VIRTIS/Rosetta imaging spectrometer. Icarus 272, 32–47. https://doi.org/10.1016/j.icarus.2016.02.028

Raponi, A., Ciarniello, M., Capaccioni, F., Mennella, V., Filacchione, G., Vinogradoff, V., Poch, O., Beck, P., Quirico, E., De Sanctis, M.C., Moroz, L.V., Kappel, D., Erard, S., Bockelée-Morvan, D., Longobardo, A., Tosi, F., Palomba, E., Combe, J.-P., Rousseau, B., Arnold, G., Carlson, R.W., Pommerol, A., Pilorget, C., Fornasier, S., Bellucci, G., Barucci, A., Mancarella, F., Formisano, M., Rinaldi, G., Istiqomah, I., Leyrat, C., 2020. Infrared detection of aliphatic organics on a cometary nucleus. Nat Astron 4, 500–505. https://doi.org/10.1038/s41550-019-0992-8

Reach, W.T., Vaubaillon, J., Lisse, C.M., Holloway, M., Rho, J., 2010. Explosion of Comet 17P/Holmes as revealed by the Spitzer Space Telescope. Icarus 208, 276–292. https://doi.org/10.1016/j.icarus.2010.01.020

Rietmeijer, F.J.M., 1993. Size distributions in two porous chondritic micrometeorites. Earth and Planetary Science Letters 117, 609–617. https://doi.org/10.1016/0012-821X(93)90106-J

Rivkin, A.S., DeMeo, F.E., 2019. How Many Hydrated NEOs Are There? Journal of Geophysical Research: Planets 124, 128–142. https://doi.org/10.1029/2018JE005584

Ross, H.P., Adler, J.E.M., Hunt, G.R., 1969. A statistical analysis of the reflectance of igneous rocks from 0.2 to 2.65 microns. Icarus 11, 46–54. https://doi.org/10.1016/0019-1035(69)90114-6

Rousseau, B., Érard, S., Beck, P., Quirico, É., Schmitt, B., Brissaud, O., Montes-Hernandez, G., Capaccioni, F., Filacchione, G., Bockelée-Morvan, D., Leyrat, C., Ciarniello, M., Raponi, A., Kappel, D., Arnold, G., Moroz, L.V., Palomba, E., Tosi, F., 2018. Laboratory simulations of the Vis-NIR spectra of comet 67P using sub-µm sized cosmochemical analogues. Icarus 306, 306–318. https://doi.org/10.1016/j.icarus.2017.10.015

Rozitis, B., Ryan, A.J., Emery, J.P., Christensen, P.R., Hamilton, V.E., Simon, A.A., Reuter, D.C., Asad, M.A., Ballouz, R.-L., Bandfield, J.L., Barnouin, O.S., Bennett, C.A., Bernacki, M., Burke, K.N., Cambioni, S., Clark, B.E., Daly, M.G., Delbo, M., DellaGiustina, D.N., Elder, C.M., Hanna, R.D., Haberle, C.W., Howell, E.S., Golish, D.R., Jawin, E.R., Kaplan, H.H., Lim, L.F., Molaro, J.L., Munoz, D.P., Nolan, M.C., Rizk, B., Siegler, M.A., Susorney, H.C.M., Walsh, K.J., Lauretta, D.S., 2020. Asteroid (101955) Bennu's weak boulders and thermally anomalous equator. Science Advances 6, eabc3699. https://doi.org/10.1126/sciadv.abc3699

Salisbury, J.W., D'Aria, D.M., Jarosewich, E., 1991. Midinfrared (2.5–13.5 µm) reflectance spectra of powdered stony meteorites. Icarus 92, 280–297.

Salisbury, J.W., Eastes, J.W., 1985. The effect of particle size and porosity on spectral contrast in the mid-infrared. Icarus 64, 586–588. https://doi.org/10.1016/0019-1035(85)90078-8

Salisbury, J.W., Wald, A., 1992. The role of volume scattering in reducing spectral contrast of reststrahlen bands in spectra of powdered minerals. Icarus 96, 121–128. https://doi.org/10.1016/0019-1035(92)90009-V

Salisbury, J.W., Wald, A., D'Aria, D.M., 1994. Thermal-infrared remote sensing and Kirchhoff's law: 1. Laboratory measurements. Journal of Geophysical Research: Solid Earth 99, 11897–11911. https://doi.org/10.1029/93JB03600

Sato, K., 1984. Reflectivity Spectra and Optical Constants of Pyrites ($FeS_2$, $CoS_2$ and $NiS_2$) between 0.2 and 4.4 eV. Journal of the Physical Society of Japan 53, 1617–1620. https://doi.org/10.1143/JPSJ.53.1617

Schröder, S.E., Mottola, S., Arnold, G., Grothues, H.-G., Jaumann, R., Keller, H.U., Michaelis, H., Bibring, J.-P., Pelivan, I., Koncz, A., Otto, K., Remetean, E., Souvannavong, F., Dolives, B., 2017. Close-up images of the final Philae landing site on comet 67P/Churyumov-Gerasimenko acquired by the ROLIS camera. Icarus 285, 263–274. https://doi.org/10.1016/j.icarus.2016.12.009

Schröder, S.E., Poch, O., Ferrari, M., Angelis, S.D., Sultana, R., Potin, S.M., Beck, P., De Sanctis, M.C., Schmitt, B., 2021. Dwarf planet (1) Ceres surface bluing due to high porosity resulting from sublimation. Nature Communications 12, 274. https://doi.org/10.1038/s41467-020-20494-5

Scott, E.R.D., Krot, A.N., 2014. 1.2 - Chondrites and Their Components, in: Holland, H.D., Turekian, K.K. (Eds.), Treatise on Geochemistry (Second Edition). Elsevier, Oxford, pp. 65–137. https://doi.org/10.1016/B978-0-08-095975-7.00104-2

Shkuratov, Y.G., 1987. Negative Polarization of Sunlight Scattered from Celestial Bodies - Interpretation of the Wavelength Dependence. Soviet Astronomy Letters 13, 182.

Shkuratov, Yu.G., Opanasenko, N.V., Kreslavsky, M.A., 1992. Polarimetric and photometric properties of the Moon: Telescopic observations and laboratory simulations: 1. The negative polarization. Icarus 95, 283–299. https://doi.org/10.1016/0019-1035(92)90044-8

Spadaccia, S., Patty, C.H.L., Capelo, H.L., Thomas, N., Pommerol, A., 2022. Negative polarization properties of regolith simulants - Systematic experimental evaluation of composition effects. A&A 665, A49. https://doi.org/10.1051/0004-6361/202243844

Sultana, Robin (2019a): VIS-NIR-MIR Reflectance spectra of mixtures of sub-µm grains of Olivine with Iron Sulfides or Anthracite, at different concentrations. SSHADE/GhoSST (OSUG Data Center). Dataset/Spectral Data. https://doi.org/10.26302/SSHADE/EXPERIMENT_OP_20230207_001

Sultana, Robin (2019b): NIR-MIR absorbance spectrum of a pellet of Olivine (sub-µm grains) mixed with KBr. SSHADE/GhoSST (OSUG Data Center). Dataset/Spectral Data. https://doi.org/10.26302/SSHADE/EXPERIMENT_OP_20230206_002

Sultana, Robin (2019c): NIR-MIR reflectance spectra of powdered Olivine at 4 different grain sizes (from 200 µm down to sub-µm). SSHADE/GhoSST (OSUG Data Center). Dataset/Spectral Data. https://doi.org/10.26302/SSHADE/EXPERIMENT_OP_20230203_001

Sultana, Robin (2019d): NIR-MIR reflectance and emissivity spectra of powdered Olivine (sub-µm grains) mixed with KBr. SSHADE/GhoSST (OSUG Data Center). Dataset/Spectral Data. https://doi.org/10.26302/SSHADE/EXPERIMENT_OP_20230206_001

Sultana, R., Poch, O., Beck, P., Schmitt, B., Quirico, E., 2021. Visible and near-infrared reflectance of hyperfine and hyperporous particulate surfaces. Icarus 357, 114141. https://doi.org/10.1016/j.icarus.2020.114141

Sultana, R., Poch, O., Beck, P., Quirico, E., Spadaccia, S., Patty, L., Pommerol, A., Maturilli, A., Helbert, J., Alemanno, G. (2023). Polarization dataset - Reflection, emission, and polarization properties of surfaces made of hyperfine grains, and implications for the nature of primitive small bodies. Zenodo. https://doi.org/10.5281/zenodo.7649167

Sumlin, B.J., Heinson, W.R., Chakrabarty, R.K., 2018. Retrieving the aerosol complex refractive index using PyMieScatt: A Mie computational package with visualization capabilities. Journal of Quantitative Spectroscopy and Radiative Transfer 205, 127–134. https://doi.org/10.1016/j.jqsrt.2017.10.012

Tedesco, E., 2004. IRAS Minor Planet Survey, IRAS-A-FPA-3-RDR-IMPS-V6. 0, NASA Planetary Data System. http://sbn. psi. edu/pds/resource/imps. html.

Tsiganis, K., Gomes, R., Morbidelli, A., Levison, H.F., 2005. Origin of the orbital architecture of the giant planets of the Solar System. Nature 435, 459–461. https://doi.org/10.1038/nature03539

Vernazza, P., Beck, P., 2017. Composition of Solar System Small Bodies, in: Elkins-Tanton, L.T., Weiss, B.P. (Eds.), Planetesimals. Cambridge University Press, Cambridge, pp. 269–297. https://doi.org/10.1017/9781316339794.013

Vernazza, P., Castillo-Rogez, J., Beck, P., Emery, J., Brunetto, R., Delbo, M., Marsset, M., Marchis, F., O. Groussin, Zanda, B., Lamy, P., Jorda, L., Mousis, O., Delsanti, A., Djouadi, Z., Dionnet, Z., Borondics, F., Carry, B., 2017. Different Origins or Different Evolutions? Decoding the Spectral Diversity Among C-type Asteroids. AJ 153, 72. https://doi.org/10.3847/1538-3881/153/2/72

Vernazza, P., Delbo, M., King, P.L., Izawa, M.R.M., Olofsson, J., Lamy, P., Cipriani, F., Binzel, R.P., Marchis, F., Merín, B., Tamanai, A., 2012. High surface porosity as the origin of emissivity features in asteroid spectra. Icarus 221, 1162–1172. https://doi.org/10.1016/j.icarus.2012.04.003

Vernazza, P., Marsset, M., Beck, P., Binzel, R.P., Birlan, M., Brunetto, R., Demeo, F.E., Djouadi, Z., Dumas, C., Merouane, S., Mousis, O., Zanda, B., 2015. Interplanetary dust particles as samples of icy asteroids. ApJ 806, 204. https://doi.org/10.1088/0004-637X/806/2/204

Wang, X., Schwan, J., Hsu, H.W., Grün, E., Horányi, M., 2016. Dust charging and transport on airless planetary bodies. Geophysical Research Letters 43, 6103–6110. https://doi.org/10.1002/2016GL069491

Yang, B., Lucey, P., Glotch, T., 2013. Are large Trojan asteroids salty? An observational, theoretical, and experimental study. Icarus 223, 359–366. https://doi.org/10.1016/j.icarus.2012.11.025

Young, C.L., Poston, M.J., Wray, J.J., Hand, K.P., Carlson, R.W., 2019. The mid-IR spectral effects of darkening agents and porosity on the silicate surface features of airless bodies. Icarus 321, 71–81. https://doi.org/10.1016/j.icarus.2018.10.032

Zeidler, S., Posch, T., Mutschke, H., Richter, H., Wehrhan, O., 2011. Near-infrared absorption properties of oxygen-rich stardust analogs: The influence of coloring metal ions. Astronomy and Astrophysics 526, 1–10. https://doi.org/10.1051/0004-6361/201015219

Zellner, B., Leake, M., Lebertre, T., Dollfus, A., 1977. Polarimetry of Meteorites and the Asteroid Albedo Scale, in: Eighth Lunar Science Conference. Lunar and Planetary Institute, Houston, p. Abstract #1358.